\documentclass[10pt,journal,twocolumn]{IEEEtran}
\usepackage{array}
\usepackage{amsmath,graphicx,epsfig,float,subfigure,times,latexsym,verbatim,mathrsfs,amssymb,textcomp,txfonts,color,cite}
\usepackage{amsfonts}
\usepackage[flushleft]{threeparttable}
\usepackage{algorithm}
\usepackage{cleveref}
\usepackage{algorithmic}
\usepackage{url}
\usepackage{bm}
\newtheorem{proposition}{Proposition}

\title{Practical Design and Implementation of Metamaterial-Enhanced Magnetic Induction Communication }

\author{Hongzhi Guo,~\IEEEmembership{Student Member,~IEEE,}
        Zhi Sun,~\IEEEmembership{Member,~IEEE,}
        and Chi Zhou

\thanks{This work was supported by the US National Science Foundation (NSF) under Grant No. 1547908. A shorter version of this paper \cite{Guo_m2ipractical} was presented at the IEEE ICC 2016.}
\thanks{Hongzhi Guo and Zhi Sun are with Department of Electrical Engineering, Chi Zhou is with Department of Industrial and Systems Engineering, State University of New York at Buffalo, Buffalo, NY 14260. E-mail: \{hongzhig, zhisun, chizhou\}@buffalo.edu.
Zhi Sun handles the correspondence of this paper.}}
\IEEEoverridecommandlockouts
\begin{document}
\maketitle
%****************************************************************************
%
\begin{abstract}
The wireless communications in complex environments, such as underground and underwater, can enable various applications in the environmental, industrial, homeland security, law enforcement, and military fields. However, conventional electromagnetic (EM) wave-based techniques do not work due to the lossy media and complicated structures. Magnetic Induction (MI) has been proved to achieve reliable communication in such environments. However, due to the small antenna size, the communication range of MI is still very limited, especially for the portable mobile devices. To this end, Metamaterial-enhanced Magnetic Induction (M$^2$I) communication has been proposed, where the theoretical results predict that it can significantly increase the data rate and range. Nevertheless, there exists a significant gap between the theoretical prediction and the practical realization of M$^2$I: the theoretical model relies on an ideal spherical metamaterial while it does not exist in nature. In this paper, a practical design is proposed by leveraging a spherical coil array to realize M$^2$I communication. The full-wave simulation is conducted to validate the design objectives. By using the spherical coil array-based M$^2$I communication, the communication range can be significantly extended, exactly as we predicted in the ideal M$^2$I model. Finally, the proposed M$^2$I communication is implemented and tested in various environments.

\end{abstract}

%****************************************************************************
% NOTE keywords are not used for conference papers so do not populate
%% them

\IEEEpeerreviewmaketitle

%****************************************************************************

%****************************************************************************
\section{Introduction}
Although the terrestrial wireless communication has been well developed and extensively utilized, its counterpart in complex environments, including underground, underwater, and confined spaces (tunnels, pipelines, and indoor space with no network infrastructures), is still in its infancy.
Wireless communication in such environments has great potential to advance the state-of-the-art technologies in environmental sustainability, homeland security, disaster relief, and military/law enforcement intelligence, such as harsh environment monitoring, miners rescue, and mitigation in nuclear plants \cite{Markham2012,Jack_MI_UG_2007,masihpour2010cooperative,silva2016design}.
When sensors or communication devices are deployed in those environments, the lossy medium and numerous obstructions can destroy their wireless connections established by most, if not all, existing wireless techniques~\cite{Sun_MI_TAP_2010,Al-Shammaa2004,kisseleff2014throughput}. One of the solutions is to utilize the long wavelength VLF frequency signal \cite{callaham1981submarine}, which requires the antenna size as large as 100 m long. However, the wireless devices, such as sensors and robots, working in the complex environments can not accommodate such large antennas or transmit enough power. As a result, wireless communication calls for a new mechanism which can enable power-efficient data transmission with portable antennas in complex environments.

Magnetic Induction (MI) is a promising solution which has been proposed to provide reliable wireless communications in underground~\cite{Sun_MI_TAP_2010} and underwater~\cite{gulbahar2012communication,Domingo2012}. Wireless signals are transmitted by leveraging the magnetic field radiated by a loop antenna rather than using the electromagnetic (EM) wave. Thus, the path loss due to the material absorption in lossy media can be effectively reduced. Also, the MI communication enjoys a stable channel since the permeability is the same for most of media in nature.
However, MI requires huge antenna to effectively transmit and receive MI signals. The existing through-the-earth (TTE) communications \cite{through-earth-communication_2010, Lockheed_MI_2012, Michael_NIOSH_TTE_2012} systems (based on MI) require coil antennas with a diameter of several meters. If the antenna size has to be reduced to centimeter level for portable device, the transmission distance would be extremely short~\cite{karlsson2004physical}. We introduced metamaterial to MI communication in~\cite{guo2015m2i}, where an ideally negative-permeability metamaterial shell is utilized to enclose a loop antenna, as shown in Fig.~\ref{fig:review1}. The mutual induction (magnetic coupling) between the MI transceivers can be significantly enhanced by matching the negative-permeability metamaterial layer with the positive-permeability environment. The theoretical results predict that, by using the metamaterial-enhanced MI (M$^2$I) communication system, a pocket-size loop antenna can achieve around 20~m communication range with acceptable data rate (Kbps).

Despite the promising prediction, it should be noted that the theoretical results in~\cite{guo2015m2i} rely on the assumption that an ideally homogeneous isotropic metamaterial spherical layer can be readily used for M$^2$I communication. Unfortunately, such material does not exist in nature. Moreover, to date, no existing solution has been proposed to realize the metamaterial sphere that can be used in the M$^2$I proposed in \cite{guo2015m2i}. Without the practical design and implementation, the M$^2$I communication technique is only a castle in the air due to the invalidated assumptions.

%To enable the wireless communications among robots, autonomous vehicles, and sensors in underground, underwater, indoor, and many other complex environments,  in~\cite{guo2015m2i} are highly desired.
%Unfortunately, such material does not exist in nature. Moreover, to date, no existing solution has been proposed to realize the metamaterial sphere layer that can be used in M$^2$I.
%Hence,

Metamaterial is composed of periodic artificial metallic or dielectric atoms. It is able to demonstrate negative permeability and permittivity, which can manipulate EM waves in extraordinary ways~\cite{pendry1999magnetism}. The periodic metamaterial components can be organized in various fashions.  For example, in~\cite{scarborough2012experimental}, metamaterial cells form a slab, which is utilized for high-efficiency Wireless Power Transfer (WPT). In \cite{xie2012proposal}, a metamaterial cylindrical shell is fabricated to improve the accuracy of Magnetic Resonance Imaging (MRI). More relevantly, \cite{pendry2006controlling} proposes to utilize spherical metamaterial to realize invisible cloaking at GHz or THz bands in theory. When it comes to implementation, spherical cloaks are simplified to cylinders due to the complexity of the 3D structure~\cite{schurig2006metamaterial}.

There are three major challenges to design and implement the M$^2$I communication: First, it is a challenge to make metamaterial spherical. Second, in M$^2$I communication we need to not only fabricate the spherical metamaterial but also precisely control its effective permeability and thickness. Since metamaterial is a kind of effective media~\cite{EMT_theory}, the effective parameters (e.g., the value of effective permeability and thickness) are hard to be accurately determined. Third, the M$^2$I communication requires the metamaterial-enhancement not only appearing near the metamaterial layer but also can be extended to further regions, i.e., the transmission distance should be much larger than the size of the antenna. It is essential for effective wireless communications in practical applications. However, existing metamaterial implementations focus on manipulating EM waves in the very close vicinity of the metamaterial.

In this paper, we realize the M$^2$I communication through the practical design and implementation. Specifically, we propose to utilize a spherical micro coil array to realize the ideal homogeneous and isotropic metamaterial spherical layer that is needed in M$^2$I communications. We uniformly place a large number of carefully-designed small coils on a spherical shell and prove that the coil array indeed demonstrates a negative permeability.
Then, we demonstrate that this shell can enhance the radiated field by the original MI antenna, which in turn increases the communication range and data rate.  In addition, the optimal configuration of the proposed spherical coil array is found and its communication performances are similar as the ideal M$^2$I predicted in~\cite{guo2015m2i}, both of which are much better than the original MI communication in underground~\cite{Sun_MI_TAP_2010}. The results are evaluated and validated by full-wave simulations and in-lab experiments. More importantly, different from the complicated EM field analyses in \cite{guo2015m2i}, we provide more intuitive understandings on the physical mechanism of M$^2$I communications through the equivalent circuit model.
Generally, the contribution of this paper can be summarized as follows: 1) a practical design based on a spherical coil array is proposed to realize the ideal M$^2$I communication; 2) we find the optimal conditions to achieve the negative permeability and the wireless communication range is significantly increased; 3) we provide both insightful design guidelines and validations by using full-wave simulations; 4) the proposed M$^2$I communication is implemented and tested in real communication system (in-house fabricated metamaterial shell and USRP software-defined radio platform).

The following of this paper is organized as follows. The related works are presented in Section II. Then, the antenna design for M$^2$I communication and equivalent circuit model are discussed in Section III. This is followed by the wireless communication performance evaluation and full-wave validation in Section IV. After that, an implementation of M$^2$I communication is demonstrated and tested in Section V. Finally, this paper is concluded in Section VI.

\section{Related Works}
%Terrestrial wireless communications are playing more and more important roles in our daily life. Their immense popularity has motivated researchers to explore wireless communications in complex environments, such as underground, underwater, etc., to enable realtime data transmission without deploying wires or other infrastructures.
 Wireless communications in complex environments can be traced back to 1920s by using the TTE communication technologies. Both giant electric and magnetic antennas are employed to overcome the high absorption loss in underground communication system for mine rescue \cite{wait1951magnetic,durkin1984electro,gibson2010channel}. The emerging wireless sensor networks provide a more efficient and economical way to collect information in complex environments \cite{Akyildiz2002}. MI communication was introduced to underground sensor networks in \cite{Sun_MI_TAP_2010} to provide stable wireless communication channels for sensors. Due to the tiny size of wireless sensors, the giant antenna is reduced to several centimeters, which makes it inefficient. Motivated by the pioneering works in \cite{Shamonina2002,5445008}, a MI waveguide formed by an array of magnetic coils is utilized to provide a low-loss path for wireless signals to extend the communication range. The existing works on MI communication in complex environments focus on two aspects. First, the antenna and wireless channel modeling, for example, the channel models in underground and underwater environments are derived based on equivalent circuits in \cite{Sun_MI_TAP_2010} and \cite{Domingo2012}, respectively. In addition, upon the channel model, MI networking protocols, connectivity, and capacity in aforementioned environments are analyzed in \cite{gulbahar2012communication,sun2011dynamic,kisseleff2015digital}. Although MI communication is becoming more and more mature, the communication range is still limited since the operating frequency falls into HF band and the antenna size is much smaller than the wavelength, which makes the antenna electrically small and inefficient. To address this challenge, existing works use either giant antenna or large numbers of relay coils to improve the received signal strength, which require more space to accommodate the antenna or more labor to deploy the relay coils. Although there are many kinds of efficient electrically small antennas (ESA), such as helix antenna and spherical resonator, they cannot be directly applied in the considered complex environments, since they are designed for terrestrial communication and leverage electromagnetic waves \cite{best2007study}. MI communication relies on the magnetic field coupling using both reactive power and real power to reduce the absorption loss and multipath fading. It has extremely small electrical size, which is very different from existing ESAs, and thus special techniques should be adopted to improve its communication performance.

\begin{table*}
\renewcommand{\arraystretch}{1.3}
\caption{Comparison of M$^2$I with existing solutions.}
\label{tab:compare}
\centering
\begin{tabular}{|p{1.65cm}| p{4.9cm}|  p{4cm} |p{5cm} |}
    \hline
      &  Conventional antenna  & MI  &  M$^2$I \\
    \hline

    Application &   terrestrial communication, e.g., cellular communication, WIFI, and wireless sensor network \cite{hui2001input,kim2010electrically} & complex environments, e.g., underground soil \cite{sun2011dynamic,kisseleff2015digital}, underwater \cite{guo2017multiple,Domingo2012}, pipeline \cite{Sun2011a}, and oil reservoir \cite{Guo2014} & complex environments (the same as MI)\\
    \hline

    Antenna type & patch, helix, electric dipole \cite{Ziolkowski_electric,hui2001input,best2007study} & magnetic loop antenna \cite{Sun_MI_TAP_2010,Lockheed_MI_2012} & metamaterial-enhanced loop antenna \cite{guo2015m2i,scarborough2012experimental,lipworth2014magnetic,Urzhumov2011}\\
    \hline

    Mechanism & electromagnetic wave & magnetic induction \cite{Sun_MI_TAP_2010,gulbahar2012communication,sun2011dynamic,kisseleff2015digital}& enhanced magnetic induction \cite{guo2015m2i}\\
    \hline
    Propagation medium   &  air & lossy medium & lossy medium \\
    \hline
%    Frequency  &  300 MHz $\sim$ 60 GHz& $\sim$50 MHz & $\sim$50 MHz \\
%    \hline

    Distance   &   up to 2 km & $\sim$10 m & $\sim$more than 20 m \\
    \hline

    Benefits & long range;
    high data rate & low propagation loss \cite{karlsson2004physical}; stable wireless channel \cite{gulbahar2012communication}; MI waveguide to extend communication range to more than 100 m \cite{Shamonina2002,5445008}& low propagation loss; stable wireless channel; longer range than MI; M$^2$I waveguide to extend communication range \cite{guo2015m2i}\\
    \hline

     Limitation &  high absorption loss in lossy medium& short range; low data rate& metamaterial need to be well designed \cite{guo2015m2i}\\

    \hline
\end{tabular}
\end{table*}

Metamaterial formed by periodical elements can enable a large number of novel applications, among which the superlens is the most relevant to this research \cite{Scarborough2012}. To enhance the magnetic coupling, the superlens have been widely adopted in WPT \cite{scarborough2012experimental,lipworth2014magnetic,Urzhumov2011,Wang2013,ranaweera2014experimental,xie2012proposal,zhang2015spatially} and MRI \cite{yeap2016metamaterial,freire2008experimental}. Both planar and curved structures are analyzed and implemented. However, different from WPT and MRI, which utilizes near field reactive power within a narrow band, MI communication works in the transition region, which employs both reactive power and real power. Also, the communication requires a relative larger bandwidth. Therefore, to extend the communication range, we have to increase the radiated power, rather than focusing on increasing power transfer efficiency in the near field. Metamaterial antennas have been proved to be a promising way to increase the efficiency of ESA. In \cite{Ziolkowski_electric}, the performance of metamaterial enhanced antenna is discussed analytically. The results show that with a metamaterial sphere, the radiated power can be significantly increased. In \cite{ziolkowski2006metamaterial,erentok2008metamaterial}, the promise of high efficiency of metamaterial-inspired antennas are validated and several types of antennas are presented and tested. Also, the metamaterial can increase antenna bandwidth or decrease antenna coupling in an array \cite{palandoken2009broadband,bait2010electromagnetic}.

Motivated by this, M$^2$I communication is introduced to complex environments in \cite{guo2015m2i} to increase the mutual coupling between transceivers and extend the communication range. Different from \cite{Ziolkowski_electric}, the effect of the lossy medium on M$^2$I antenna is analyzed and simulated. Although an initial prototype is provided in \cite{guo2015m2i}, the size of the prototype is large and an analytical analysis to connect the ideal EM analysis with the prototype is missing. Also, there is a lack of practical implementation guideline on M$^2$I communication to achieve the promising results in \cite{guo2015m2i}. In this paper, we propose a spherical coil array to realize the metamaterial sphere and M$^2$I communication. Different from \cite{guo2015m2i}, which is based on EM analysis and simulation, this paper fills the gap between ideal model and real implementation by using equivalent circuit analysis to provide more intuitive understandings. There are various approaches to fabricate metamaterials in 2D and 3D \cite{zedler2008systematic}. We adopt a 3D spherical structure with coil arrays to realize the proposed M$^2$I communication. The unique characteristics of MI and M$^2$I are also summarized in Table~\ref{tab:compare}.

\section{Antenna Design for Practical M$^2$I Communication}
In this section, we first review the ideal M$^2$I communication in \cite{guo2015m2i} and extract the key antenna and metamaterial design objectives. After that, we introduce the spherical coil array-enabled metamaterial shell to realize the design objectives. Meanwhile, the effective parameters and resonance conditions are discussed.

\subsection{Design Objectives}

\begin{figure}[t]
  \centering
    \includegraphics[width=0.22\textwidth]{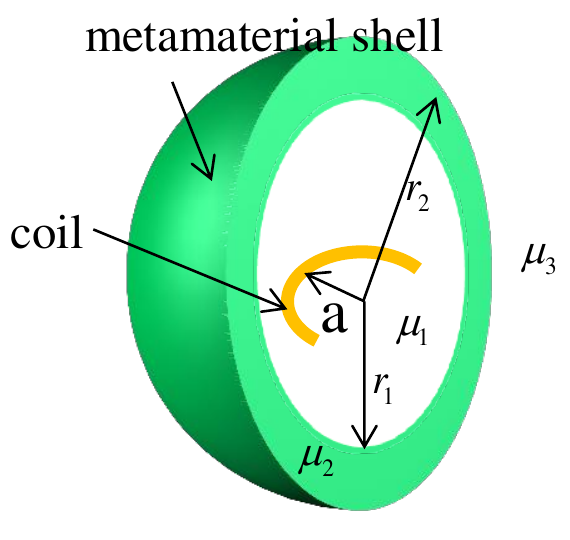}
    \vspace{-5pt}
  \caption{Illustration of ideal M$^2$I.}
  \vspace{-15pt}
  \label{fig:review1}
\end{figure}

The geometric structure of the antenna for M$^2$I communication in \cite{guo2015m2i} consists of two part, namely, the active loop antenna in the center and a passive metamaterial shell with inner radius $r_1$ and outer radius $r_2$, as shown in Fig.~\ref{fig:review1}. The permeability of the inner layer, metamaterial layer and outer layer are denoted by $\mu_1$, $\mu_2$ and $\mu_3$, respectively. Also, the metamaterial layer in \cite{guo2015m2i} is considered as homogeneous and isotropic. The key parameter of MI communication is the mutual inductance, which couples the transmitting and receiving antenna together. By using M$^2$I, the mutual inductance can be expressed as $M_0={\mathcal F_1}/{\bar S_{m}^2}$, where
\begin{align}
\label{equ:sm}
\textstyle
{\bar S_{m}}={\mathcal F_2}{\left[2r_1^3(\mu_1-\mu_2)(\mu_3-\mu_2)-r_2^3(2\mu_2+\mu_1)(2\mu_3+\mu_2)\right]}+{\hat o},
\end{align}
${\mathcal F_1}$ and ${\mathcal F_2}$ are coefficients, and ${\hat o}$ is an asymptotically small value. The first term on the right-hand side of \eqref{equ:sm} is much larger than ${\hat o}$. When the term in the bracket of \eqref{equ:sm} is zero, ${\bar S_m}$ can be minimized, which in turn maximizes $M_0$. Therefore, if the outer shell radius $r_2$, inner permeability $\mu_1$ and outer permeability $\mu_3$ are determined, by adjusting metamaterial shell's thickness $r_1$ and permeability $\mu_2$, we can always maximize $M_0$.

This enhancement is because of the matching between the metamaterial layer and the antenna \cite{Ziolkowski_electric}. In view of \eqref{equ:sm}, only a negative $\mu_2$ can make the dominant part vanish. Hence, the first key design objective is a negative-permeability layer. Meanwhile, this layer should be spherical and thus the second objective is designing a spherical metamaterial. Last but not least, the resonance condition of the spherical shell should be found upon rigorous analyses. With these objectives in mind, we provide a practical design in the following.
%\subsection{Negative Effective Permeability and Spherical Shell}
\subsection{Metamaterial Modeling and Analysis}
\label{sec:negativemu}

\begin{figure}[t]
  \centering
    \includegraphics[width=0.25\textwidth]{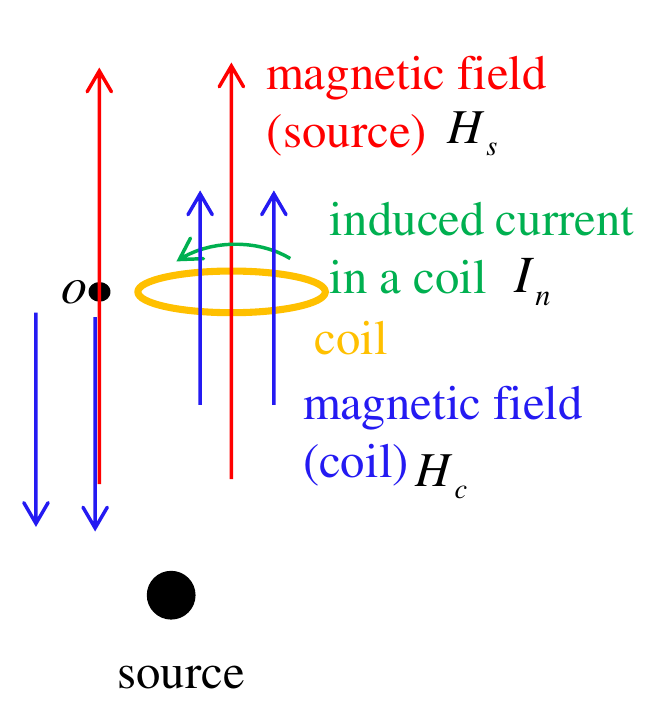}
  \caption{ Illustration of effective $\mu$. An ideal source generates magnetic field $H_s$ (red color), which induces current $I_n$ in a coil. The current direction can be controlled by varying its reactance. The coil reradiates magnetic field $H_c$ (blue color).}
    \label{fig:ideal2}
    \vspace{-15pt}
\end{figure}

When the particles in a mixture are much smaller than the wavelength, the mixture can be regarded as an effective homogeneous medium and its constitutive parameters, such as effective permeability, permittivity and conductivity, can be derived analytically \cite{EMT_theory}. Thus, to design the metamaterial, we need large numbers of atoms and their size is supposed to be much smaller than wavelength. Before demonstrating how to find the effective parameter, we give an example by using a coil to show how it can change the effective permeability.

As shown in Fig. \ref{fig:ideal2}, a source radiates magnetic field $H_s {\hat s}$, which can induce current $I_n$ in a coil (the coil has a capacitor in series to tune its reactance). The direction of $I_n$ can follow right-hand rule or left-hand rule, which is determined by its reactance. Here, ${\hat s}$ is a unit vector standing for the direction of magnetic field. If we consider there is no coil, at the observation point $o$ in the figure, the magnetic flux can be expressed as $B_m =\mu_0 H_s {\hat s}$, where $\mu_0$ is the permeability of vacuum. However, if the coil exists, due to the induced current $I_n$, the coil reradiates magnetic field $H_c {\hat c}$, where ${\hat c}$ is also a unit vector denoting the direction of the reradiated magnetic field. As a result, the magnetic flux at point $o$ can be updated as
\begin{align}
\label{equ:effectivemu}
B_m=\mu_0 H_s {\hat s}+\mu_0 H_c {\hat c}=\mu_0\left(1+\frac{H_c {\hat c}}{H_s {\hat s}}\right)H_s {\hat s}=\mu_{eff}H_s {\hat s},
\end{align}
where $\mu_{eff}$ is the effective permeability at point $o$.

In view of \eqref{equ:effectivemu}, the effective permeability can be controlled provided that the reradiated magnetic field $H_c{\hat c}$ can be manipulated. When ${H_c {\hat c}}/{H_s {\hat s}}$ is negative and its absolute value is larger than 1, the negative permeability can be obtained. The detailed approach to adjust $H_c{\hat c}$ is discussed in the following sections. Also, it is worth noting that besides coil, there are many other radiators can be utilized to achieve negative permeability, such as complicated metallic structures and dielectric spheres \cite{caloz2005electromagnetic}. Since coil is relatively easy to fabricate and its performance is tractable, we employ it in this paper. Next, more coils are utilized to construct an effective medium with negative permeability.

\begin{figure}[t]
\begin{minipage}{0.24\textwidth}
\centering
  \includegraphics[width=0.9\textwidth]{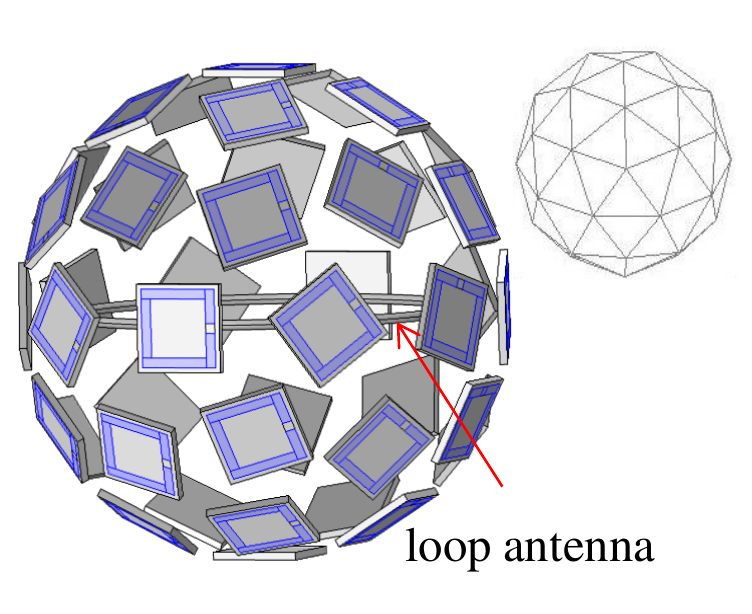}
  \caption{Spherical coil array enclosed loop antenna. }
    \label{fig:geo1}
\end{minipage}\quad
\begin{minipage}{0.19\textwidth}
\centering
  \includegraphics[width=0.9\textwidth]{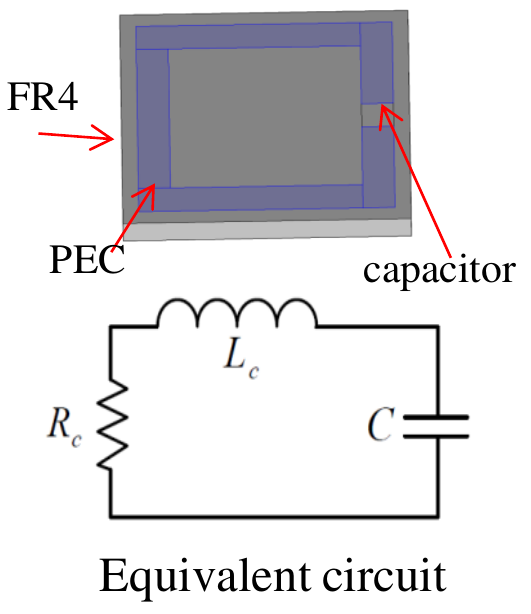}
  \caption{A coil unit on the shell. }
    \label{fig:geo2}
\end{minipage}
\vspace{-12pt}
\end{figure}

The geometrical structure of the metamaterial is presented in Fig. \ref{fig:geo1}. The same as our discussion in \cite{guo2015m2i}, we set the outer radius of the spherical shell as 0.05~m, which can be accommodated by wireless sensors/robots. By using the Pentakis icosidodecahedron \cite{horn1984extended}, we equally divide the surface of the sphere into 80 triangles with 42 vertices. Each vertex is the center of a coil and the coil's orientation is the radial direction from the center of the shell to the vertex. In this way, the coil array is constructed by 42 identical coils. The coils are made of copper with capacitors to tune them and the simulation model of a PCB unit and corresponding equivalent circuit are depicted in Fig.~\ref{fig:geo2}. In the equivalent circuit $L_c$, $R_c$, and $C$ represent the coil inductance, resistance, and compensation capacitance, respectively. The square coil employed here has similar performance as a circular coil since it is much smaller than the wavelength. In the following, we show that this structure can achieve negative permeability and it is equivalent to the ideal metamaterial shell in \cite{guo2015m2i}.

\begin{figure}[t]
  \centering
    \includegraphics[width=0.3\textwidth]{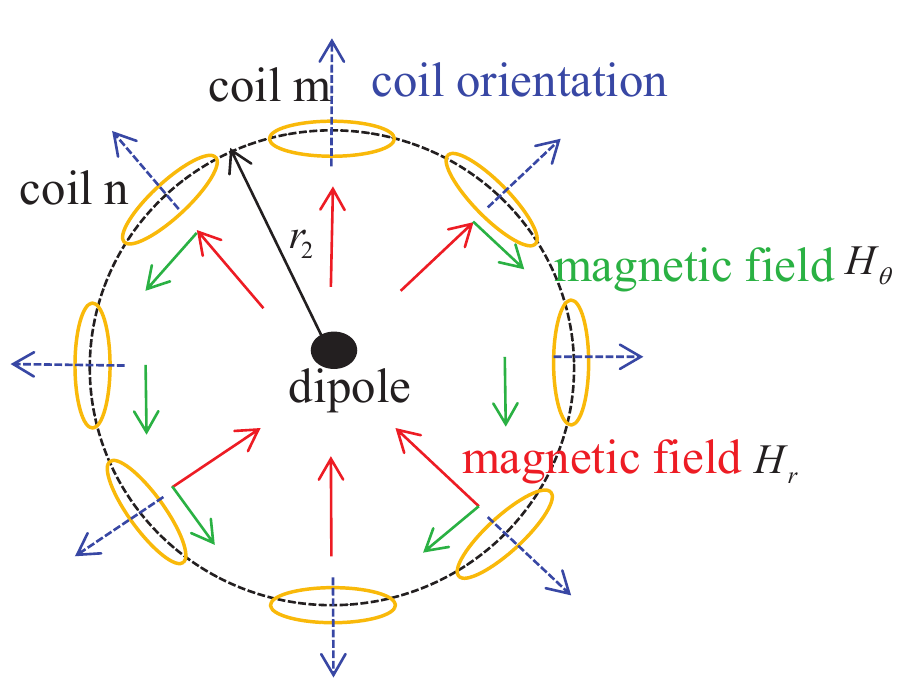}
    \vspace{-5pt}
  \caption{Illustration of coil orientation and the direction of magnetic field radiated by a loop antenna. }
  \vspace{-15pt}
  \label{fig:ideal12}
\end{figure}
First, the spherical coil array is considered as a homogeneous layer. By using effective medium theory \cite{pendry1999magnetism}, we find the effective permeability of this layer. The radiation source is the same as our work in \cite{guo2015m2i}, i.e., a magnetic loop antenna, and all the coils on the shell are passive, as shown in Fig. \ref{fig:ideal12}. According to Kirchhoff voltage law we can obtain,
\begin{align}
\label{equ:app1}
{\bf Z}\cdot {\bf I}={\bf V}
\end{align}
where ${\bf Z} \in \mathbb{C}^{N\times N}$,
$N$ is the number of coils on the shell, the diagonal elements $Z_n=R_c+j\omega L_c+1/j\omega C$, $Z_{mn}=j \omega M_{mn}$, $M_{mn}$ is the mutual inductance between coil $m$ and coil $n$ and the detailed calculation is provided in Appendix A, ${\bf I}\in \mathbb{C}^{N\times 1}$ and its element $I_n$ is the induced current in coil $n$, and ${\bf V}\in \mathbb{C}^{N\times 1}$ and its element $V_n$ is the voltage. The magnetic field radiated by a coil or loop antenna in radial direction $H_r$ and polar direction $H_{\theta}$ can be found in Appendix A. For a planar spiral coil on a PCB, its self-inductance can be approximated by \cite{mohan1999simple}
\begin{align}
L_c\approx\frac{1.27\mu_0 d_{avg}}{2}\left[\ln\left(\frac{2.07}{\rho}\right)+0.18\rho+0.13\rho^2\right],
\end{align}
where $d_{avg}=0.5(d_i+d_o)$, $\rho={(d_o-d_i)}/{(d_i+d_o)}$, $d_o$ is the outer edge length of the square coil and $d_i$ is the inner length. The coil's DC resistance is $R_d=\rho_c l_c/(t_w t_c)$ where $\rho_c$, $l_c$, $t_w$ and $t_c$ are the resistivity, length, width, and thickness of the trace, respectively. The trace's skin depth is $\delta_c=\sqrt{\rho_c/(\pi\mu_0 f)}$. Since the operating frequency of M$^2$I antenna is above 1~MHz, the AC resistance should be considered which is \cite{jow2007design,wheeler1942formulas}
\begin{align}
R_c=\frac{R_d t_c}{\delta_c(1-e^{-t_c/\delta_c})}.
\end{align}
The series capacitance can be chosen as wish and thus we can carefully design it to achieve optimal performance.

The direction of $H_r$ and $H_{\theta}$ are depicted in Fig. \ref{fig:ideal12}. Since the coil orientation is parallel with $H_r$'s direction and perpendicular to $H_{\theta}$'s direction, only $H_r$ can induce currents in the coils on the shell. As a result, this shell can be regarded as metamaterial for $H_r$, but not for $H_{\theta}$. This can be solved by using a tri-directional coil \cite{Guo_underwater} or a cubic metamaterial as that in \cite{scarborough2012experimental,lipworth2014magnetic}. In this paper, we only consider $H_r$ to derive succinct analytical results to extract more insightful understandings. Consequently, $V_n$ in \eqref{equ:app1} can be expressed as

\begin{align}
\label{equ:voltage}
V_n=-j\omega\pi a^2 \mu_0 {H_r} \cos{\theta_n},
\end{align}
where $\theta_n$ is the angle of $H_r$ and the coil's orientation. According to Kirchhoff's voltage law, without loss of generality, in coil $n$ we can obtain
%\begin{align}
%\label{equ:kirchhoff}
%I_n(R_c+j\omega L_c-\frac{j}{\omega C})+\sum_{i=1,i\neq n}^{N}j\omega M_{in}I_i=-j\omega\pi a^2 \mu_0 { H_r} \cos \theta_n.
%\end{align}
%Rearranging \eqref{equ:kirchhoff} gives,
\begin{align}
\label{equ:rearrange1}
\frac{I_n(R_c+j\omega L_c-\frac{j}{\omega C})}{{H_r \cos \theta_n}}=-j\omega\pi a^2 \left(\sum_{i=1,i\neq n}^{N}\frac{M_{in}I_i}{\pi a^2 {H_r \cos\theta_n}}+\mu_0\right).
\end{align}
In addition, we have the following proposition:
\begin{proposition}
\label{pro:IH}
For any coil $n$ on the shell, $I_n={\mathcal A} {H_r} \cos \theta_n$, where ${\mathcal A}$ is a constant.
\end{proposition}
The proof is provided in Appendix B. An intuitive understanding of Proposition~\ref{pro:IH} is that the induced current in a coil on the sphere is proportional to the incoming magnetic field radiated by the loop antenna.

If only coil $n$ exists on the spherical shell and all other coils are removed, only $\mu_0$ is in the parentheses on the right-hand side of \eqref{equ:rearrange1}. In other words, the first term in the parentheses denotes the mutual interactions among coils on the shell. When there are no other coils, this term does not exist and this is indeed the conventional material with permeability $\mu_0$. Hence, when we consider all other coils on the shell, the terms in the parentheses can be regarded as the effective permeability. In addition, based on Proposition~\ref{pro:IH}, since ${I_n}/({{ H_r} \cos \theta_n})={\mathcal A}$, the left-hand side of \eqref{equ:rearrange1} is a constant. Therefore, no matter which coil on the shell we select, the value in the parentheses does not change, i.e., the effective permeability is homogeneous.

By using \eqref{equ:rearrange1} and the value of ${\mathcal A}$ in Appendix B, we can obtain
\begin{align}
\label{equ:effmu}
\mu_{eff}=\left(1-\frac{j\omega \sum_{i=1,i\neq n}^{N}M_{in}}{R_c+j\omega L_c+1/j\omega C+\sum_{i=1,i\neq n}^{N}j\omega M_{in}}\right)\mu_0.
\end{align}
From \eqref{equ:effmu} we can see when the second term in the parentheses is larger than 1, a negative permeability can be achieved. Moreover, from \eqref{equ:effmu} we can find that the effective permeability is only determined by the coils' mutual inductances and the lumped elements. It is not affected by the radiation source. So far, we have addressed the first two challenges to design a spherical negative-permeability metamaterial. Next, we show the way to change metamaterial effective permeability and find the optimal parameters.

\subsection{Optimal Design of M$^2$I Antenna}
In view of \eqref{equ:effmu}, the second term in the parentheses need to be larger than 1 and thus a negative permeability can be achieved. To distinguish the real and imaginary part, \eqref{equ:effmu} can be written as,
\begin{align}
\label{equ:effmu2}
\mu_{eff}=\left[1-\frac{\omega \sum_{i=1,i\neq n}^{N}M_{in}X}{R_c^2+X^2}-j\frac{\omega \sum_{i=1,i\neq n}^{N}M_{in}R_c}{R_c^2+X^2}\right]\mu_0,
\end{align}
where $X=\omega L_c-\frac{1}{\omega C}+\omega \sum_{i=1,i\neq n}^{N}M_{in}$. In order to obtain negative permeability, $F=\frac{\omega \sum_{i=1,i\neq n}^{N}M_{in}X}{R_c^2+X^2}$ need to be maximized to let it at least be larger than 1. Then, we find the derivative of $F$ with regard to $X$ and the optimal condition is $X^2=R_c^2$. Hence, $X$ can be positive $R_c$ or negative $R_c$. The term $\sum_{i=1,i\neq n}^{N}M_{in}$ is negative and thus when $X=R_c$ a large positive $\Re(\mu_{eff})$ can be obtained, while $X=-R_c$ a large negative $\Re(\mu_{eff})$ can be obtained, which is $1+({\omega \sum_{i=1,i\neq n}^{N}M_{in}})/({2R_c})$. Since our objective is to design negative permeability metamaterial, $X=-R_c$ is the prerequisite, upon which we can obtain the resonance frequency, which is
\begin{align}
\label{equ:solution}
\omega=\frac{-R_c\pm \sqrt{R_c^2+4(L_c+\sum_{i=1,i\neq n}^{N}M_{in})/C}}{2(L_c+\sum_{i=1,i\neq n}^{N}M_{in})}.
\end{align}
Observe that, in \eqref{equ:solution} there is a meaningful positive solution and a meaningless negative solution. Consequently, there is only one resonant frequency, which results in large negative permeability. Note that, besides frequency, the capacitance $C$ is also a variable we can freely change. Equivalently, if $\omega_0$ is the designed operating frequency, by varying $C$ we can also achieve the resonance and the optimal value is
\begin{align}
\label{equ:capacitance_res}
C=\frac{1}{\omega_0(R_c+\omega_0 L_c+\omega_0\sum_{i=1,i\neq n}^{N}M_{in})}.
\end{align}

The above optimal condition can only predict the most negative permeability, which may not be the optimal condition of M$^2$I communication, because when the real part is significant, this also yields a large imaginary part. However, since the negative permeability is based on coil unit's resonance, it can only be achieved within very narrow band or small range of capacitance variation. Thus, the aforementioned optimal condition can be a good approximation. By examining \eqref{equ:effmu2} we can find that the imaginary part of the effective permeability is mainly determined by $R_c$. Specifically, if we ideally assume that $R_c=0$, the imaginary part vanishes and by changing $X$ or, equivalently, changing $C$, we can obtain any $\mu_{eff}$ as wish. As a result, the bottleneck becomes the coil's resistance $R_c$ and it needs to be minimized to reduce the negative effects. To reduce $R_c$, we may aggressively increase the width of the trace, use lower frequency to minimize AC resistance, or reduce the trace's length. Nevertheless, these solutions are either not practical or not efficient. Here, we show that by increasing the operating frequency we can obtain more negative $\mu_{eff}$, since $\omega$ increases faster than $R_c$ at lower MHz band. In other words, use higher frequency, we can increase the ratio ${\omega \sum_{i=1,i\neq n}^{N}M_{in}}/({2R_c})$.

\begin{figure}[t]
  \centering
  \subfigure[$f_0$ is 20~MHz]{
    \label{fig:mu_eff_f_20MHz1}
    \includegraphics[width=0.22\textwidth]{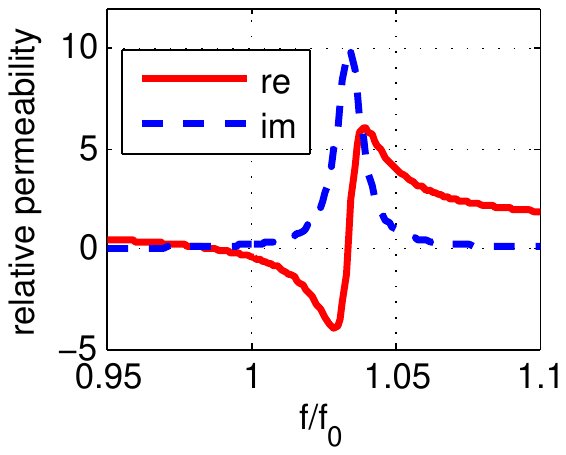}}\quad
  \subfigure[$f_0$ is 50~MHz]{%
    \includegraphics[width=0.22\textwidth]{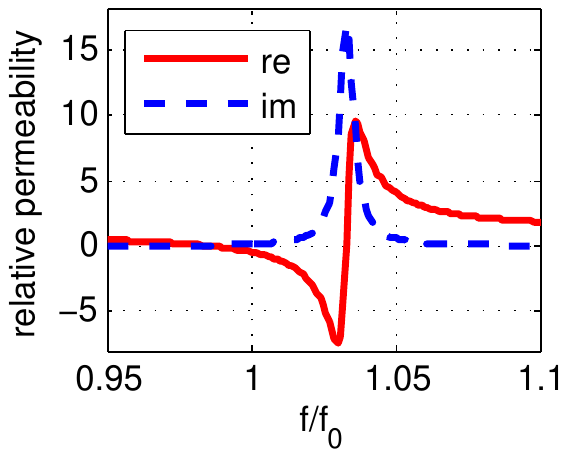}\quad
    \label{fig:mu_eff_f_50MHz}}
      \vspace{-5pt}
  \caption{Effect of frequency on the effective permeability of the coil array. }
    \vspace{-15pt}
  \label{fig:mu_eff_f}
\end{figure}

\begin{figure}[t]
  \centering
  \subfigure[20~MHz]{
    \label{fig:20MHz_eff_mu}
    \includegraphics[width=0.22\textwidth]{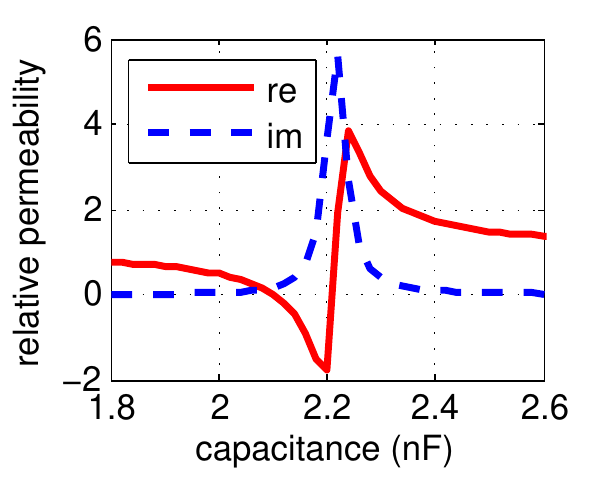}}\quad
  \subfigure[50~MHz]{%
    \includegraphics[width=0.22\textwidth]{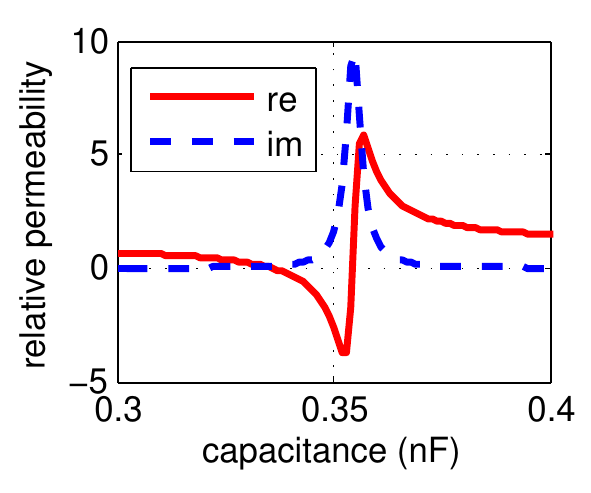}\quad
    \label{fig:mu_eff_50MHz}}
      \vspace{-5pt}
  \caption{Effect of capacitance on the effective permeability of the coil array. }
    \vspace{-20pt}
  \label{fig:mu_eff}
\end{figure}

Based on the developed model, we analyze the effective permeability numerically. First, the configurations of the small coil are given as: $d_o$=16~mm, $d_i$=12~mm, and $t_c$=0.0356~mm. The small coil is made as large as possible to increase the filling ratio of the metamaterial. The trace is made of copper and its permeability is $\mu_0$=$4\pi\times10^{-7}~$H/m. The analytical results in \eqref{equ:solution} and \eqref{equ:capacitance_res} indicate that either varying frequency or capacitance, effectively negative permeability can be obtained. Based on this observation, without loss of generality, we first fix the capacitance as 2.16~nF and 0.35~nF to investigate the resonance frequency, then fix the frequency as 20~MHz and 50~MHz to investigate the effect of capacitance. The frequency effect is shown in Fig. \ref{fig:mu_eff_f}. As predicted, there is a resonance as the frequency varies. It is worth noting that the negative effective permeability appears at lower frequency than the large positive permeability. It can be interpreted from two aspects. First, only around the resonance frequency the small coils on the sphere can respond efficiently no matter positive or negative due to the small impedance in the coil, i.e., either significantly large or small frequency cannot excite the coils because of high impedance. Second, when the frequency is a little smaller than the optimal frequency, the impedance of the capacitor is large since $Z_c=\frac{1}{j\omega C}$ and the overall reactance of the small coil is negative. According to right-hand rule, a coil with positive inductance generates a current, which radiates magnetic field whose direction is opposite to the incoming magnetic field. However, now due to the contribution of the capacitor, the negative reactance leads to a current which radiates magnetic field whose direction is the same as the incoming magnetic field. In other words, left-hand rule governs the EM field in this scenario. Therefore, the effective permeability becomes negative. When the frequency becomes a little larger than the optimal frequency, the impedance of the capacitor becomes smaller due to the increase of $\omega$ and the overall reactance of the coil returns to positive which obeys the right-hand rule again. Thus, the effective permeability is positive.

Next, we gradually change the capacitance and fix the operating frequency at 20~MHz and 50~MHz, as shown in Fig.~\ref{fig:mu_eff}, resonances can also be obtained. The physical principle is the same as varying the frequency since both the frequency and capacitance are inversely proportional to the capacitor's impedance. By using the optimal condition derived in \eqref{equ:solution} and \eqref{equ:capacitance_res}, we find the optimal frequency for 2.16~nF is 20.17~MHz and the resonance capacitance for 20~MHz is 2.19~nF, which agree well with the results shown in Fig.~\ref{fig:mu_eff_f_20MHz1} and Fig.~\ref{fig:20MHz_eff_mu}. In the same way, we find the optimal frequency for 0.35~nF is 50.48~MHz and the resonance capacitance for 50~MHz is 0.353~nF, which also confirm the results in Fig.~\ref{fig:mu_eff_f_50MHz} and Fig.~\ref{fig:mu_eff_50MHz}. Thus, \eqref{equ:solution} and \eqref{equ:capacitance_res} can predict the optimal frequency and capacitance with good accuracy.

\section{Wireless Communication Performance Analysis}
The spherical metamaterial shell with negative permeability has been designed in preceding section, upon which we analyze its wireless communication performance in this section. Moreover, the high efficiency of M$^2$I communication is explained from the perspective of equivalent circuit model rather than the complicated EM field model in \cite{guo2015m2i} and this can provide more straightforward understanding of M$^2$I communication. Additionally, the predicted negative self-inductance by the ideal model in \cite{guo2015m2i} is explained here.

\subsection{Magnetic Field Enhancement Analysis}
\label{sec:rfa}
In order to compare the efficiency of M$^2$I communication and MI communication, we adopted a similar approach as that in \cite{Ziolkowski_electric} which provides the transmitting antennas with the same input current 1~A and analytically calculate and numerically simulate the magnetic field intensity radiated by the antennas.
Note that, the radius of the M$^2$I loop antenna is set as 0.04~m, while the radius of the MI loop antenna is set as 0.05~m, which is also the outer radius of the M$^2$I shell. Moreover, the ideal M$^2$I's performance is not comparable with coil array-based M$^2$I antenna, although they have the same negative permeability. Because the inner radius of ideal M$^2$I in \cite{guo2015m2i} is 0.025~m and the loop antenna is 0.015~m in radius. However, the coil array-based M$^2$I antenna has almost no physical thickness. As a result, we can set the loop antenna much larger than ideal M$^2$I antenna to fully use the space.
\subsubsection{Magnetic Field Analysis}

As depicted in Fig. \ref{fig:geo1}, in M$^2$I communication only the large loop antenna is actively excited and all other small coils on the shell are passive.
Since the induced voltage in the receiving coil is proportional to magnetic field intensity, we use it as a metric to show the enhancement. For MI antenna, the magnetic fields are radiated only by the loop antenna. However, for M$^2$I antenna, the magnetic field is the vector superposition of that radiated by the loop antenna inside the shell and the reradiated magnetic fields by the passive coils, which can be expressed as
\begin{align}
\label{equ:field}
{\hat H}=\sum_{i=0}^{N}\left[H_r(I_i){\hat r_i}+H_{\theta}(I_i){\hat \theta_i}\right],
\end{align}
where the large loop antenna is denoted by $i=0$. To find $I_i$, we can use \eqref{equ:app1} and \eqref{equ:voltage}.
Note that to find $H_r$ and $H_{\theta}$ the expressions in Appendix A can be applied. However, the formulas are valid when the coordinates origin is the center of the coil and z-axis overlaps the coil's orientation vector. Otherwise, we have to do transformation to convert all the fields in the same coordinates. To find the magnetic field intensity at an observation point, we first consider all the fields from each antenna individually. Then we decompose the fields in a Cartesian coordinates. Finally, we add all the fields together to obtain the overall magnetic field at the point.

\begin{figure}[t]
  \centering
    \includegraphics[width=0.4\textwidth]{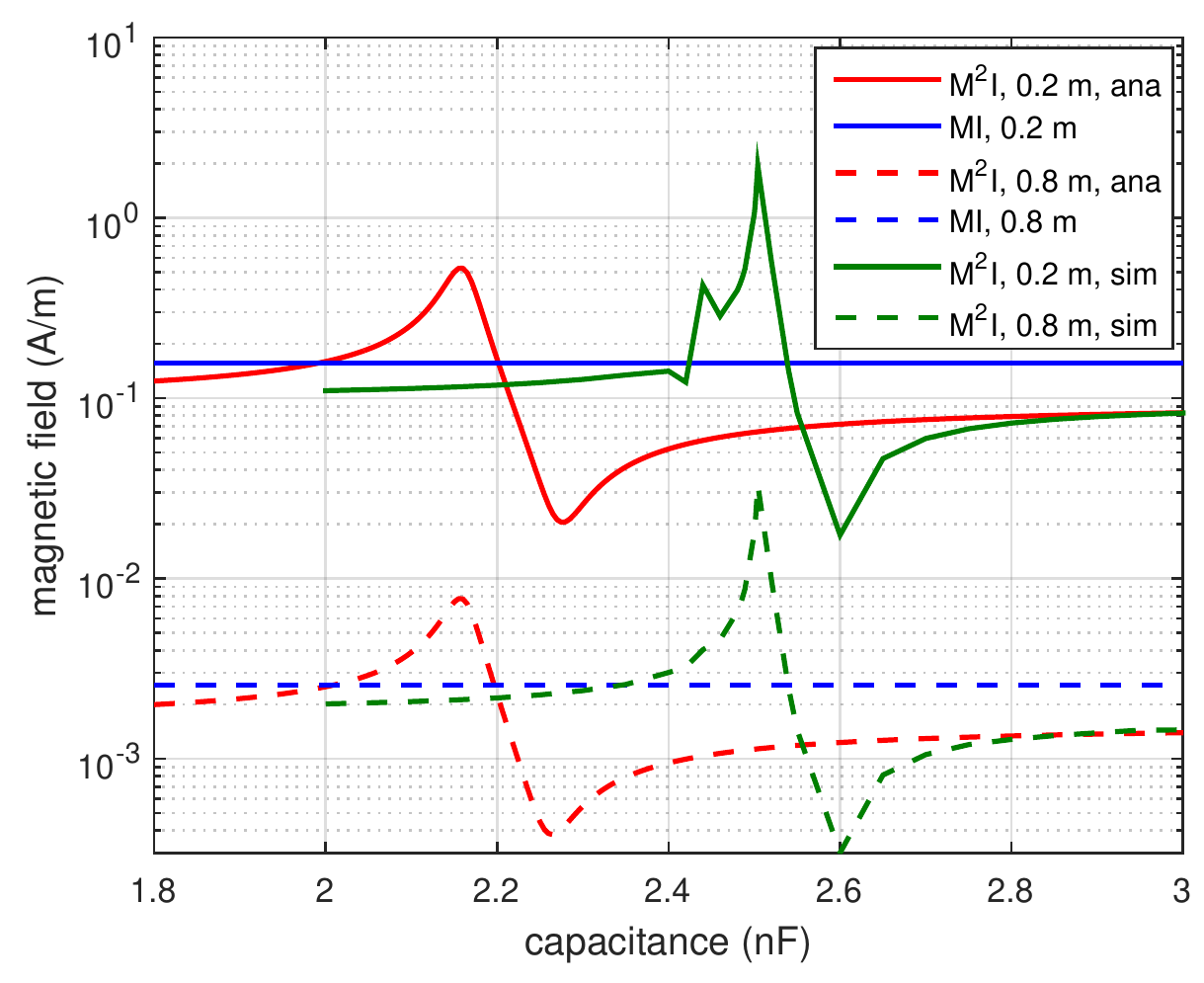}
    \vspace{-5pt}
  \caption{Magnetic filed intensity when operating frequency is 20~MHz. }
  \vspace{-15pt}
  \label{fig:radiation_20MHz}
\end{figure}

\begin{figure}[t]
  \centering
    \includegraphics[width=0.4\textwidth]{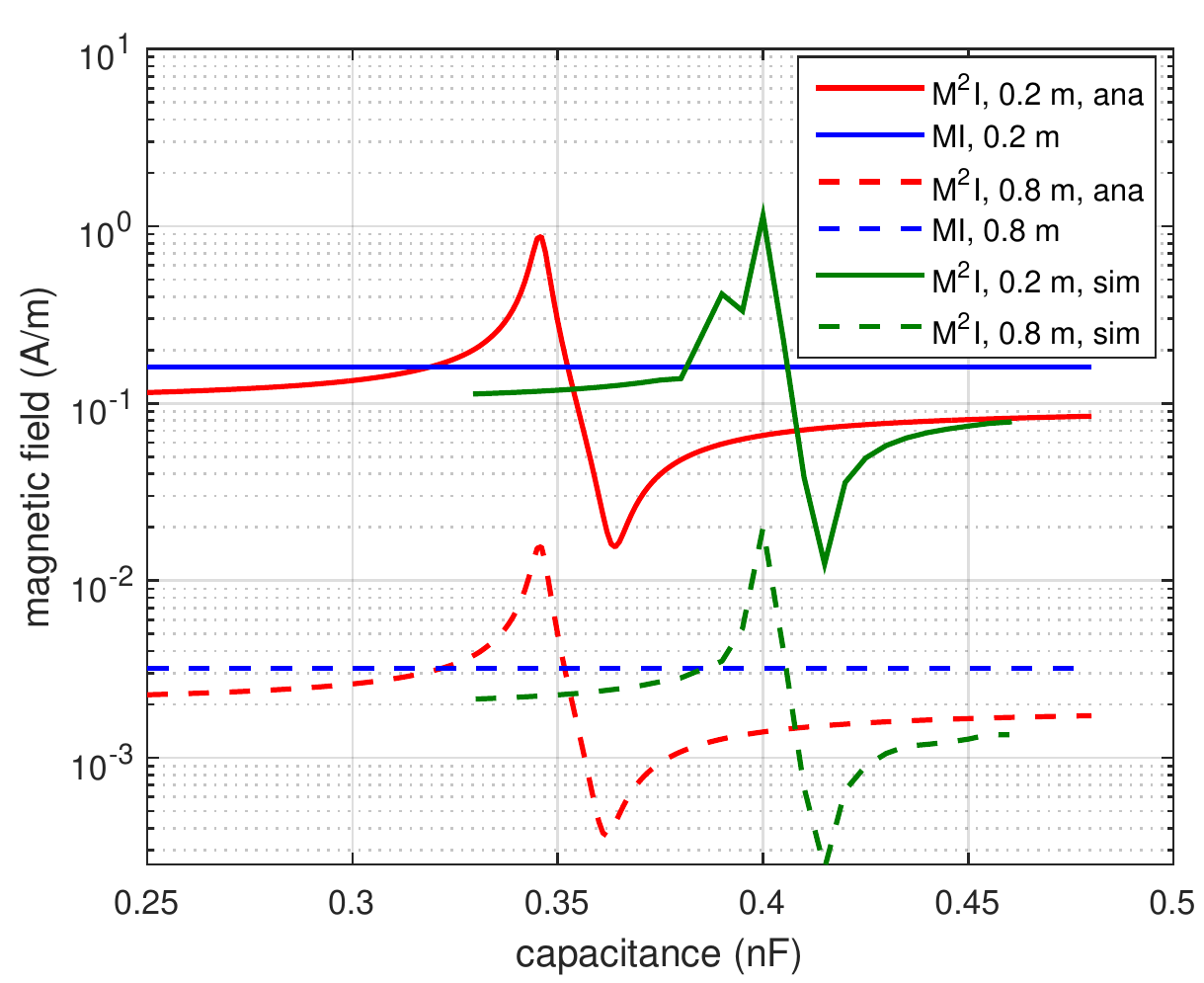}
    \vspace{-5pt}
  \caption{Magnetic filed intensity when operating frequency is 50~MHz. }
  \vspace{-15pt}
  \label{fig:radiation_50MHz}
\end{figure}

Based on the developed analytical model and full-wave simulation, we analytically and numerically evaluated the performance of M$^2$I communication. We still consider 20~MHz and 50~MHz as the operating frequency and keep the distance as 0.2~m and 0.8~m to show the gain is stable with distance. As demonstrated in preceding section, either varying capacitance or frequency, we can obtain the resonance. Here, we keep the frequency as a constant and vary the capacitance. The results are plotted in Fig.~\ref{fig:radiation_20MHz} and Fig.~\ref{fig:radiation_50MHz}. However, observe that there is a deviation of the resonance capacitance between the analytical results and the simulation results. This is due to the approximation of mutual inductance and self-inductance which are not exactly the same as those in the full-wave simulation model and thus the resonance capacitor shifts a little bit. Nevertheless, the curve shape of the analytical results match well with the simulation results which also reveal the following two facts. First, the resonance does not appear when the effective permeability is the most negative. As shown in the two figures, the resonance is achieved when the real permeability is around -1 rather than even more negative values. On the one hand, as will be shown in following discussions, negative permeability can enhance the radiated magnetic field. On the other hand, the more negative, the higher effective loss which reduces the shell's efficiency. Therefore, there exists an optimal permeability value. Second, the 50~MHz achieves even better performance than 20~MHz in the analytical results, while their simulation results are similar. The difference between the analytical model and the simulation model is the coil resistance. In the analytical model, the real PCB trace has a certain resistance. However, in the simulation model, the trace is perfect electric conductor, which has almost no resistance. As a result, the performance of the simulation model is nearly ideal since it has almost no loss, while the performance of the analytical model highly depends on the ratio ${\omega \sum_{i=1,i\neq n}^{N}M_{in}}/({2R_c})$ as discussed in preceding section. $R_c$ is determined by the AC resistance. From 20~MHz to 50~MHz, $\omega$ increases faster than $R_c$, and thus 50~MHz performs better. We can foresee that as frequency increases even more, when $R_c$ increases faster than $\omega$, high frequency cannot achieve better performance.

\subsubsection{Physical Principle of Enhanced M$^2$I Communication}
\label{sec:reradiated}

As discussed in the design objectives, only if the dominant part on the right-hand side of \eqref{equ:sm} becomes zero, the radiated magnetic field can be enhanced and significantly large mutual inductance can be achieved. Accordingly, we have to use metamaterial to make $\mu_2$ negative. Now, from a different perspective, we consider the shell consists of discrete coils and show how each coil works to let the shell resonant. Also, we provides insights to understand the peaks and nulls in Fig.~\ref{fig:radiation_20MHz} and Fig.~\ref{fig:radiation_50MHz}.

As discussed in Section \ref{sec:negativemu}, by manipulating the reradiated magnetic field by coils we can change the effective permeability. Similarly, we can adjust the capacitance to change the direction of the reradiated magnetic field to either enhance or cancel the radiated field by the loop antenna. To illustrate the physics better, we assume that the small coil's resistance $R_c$ is infinitely small which is equivalent to a no-loss metamaterial unit, and its reactance $X=\omega L_c-1/(\omega C)+\omega \sum_{i=1,i\neq n}^{N}M_{in}$ is either positive or negative. Then, assume that the magnetic field $H_s$ radiated by the loop antenna induces a voltage $V=-j\omega \pi a^2 \mu_0 H_s=-jV_0$ in a small coil. Now we consider three scenarios to distinguish the resonance and nonresonance, and the peaks and nulls.

First, when the small coil is far from resonance, i.e., $X$ is dramatically large, although the induced voltage is large, the induced current in the small coils are extremely small. As a result, the effect of those small coils on the shell can be neglected and only the loop antenna in the center is radiating. Therefore, under this circumstance, M$^2$I has no gain over MI and, even worse, its efficiency is lower than MI since its radiating loop antenna is within a shell which is smaller than MI's antenna. That is why on the two sides of the resonance, M$^2$I's field intensity is lower than MI. Then, let us look at the resonance region. First, at the peak point, $X<0$ since $C$ is smaller than the capacitance that can make the reactance vanish. When $R_c$ is negligible, the current in each small coil is $V_0/|X|$, which has the same direction as the current in the radiating loop antenna. Therefore, the coils on the shell can collect the near field power of the radiating loop antenna and reradiate this out. This phenomenon is strong because the small coil is close to resonance and $X$ is small, which results in a large current in the coil. Thus, the reradiated field is significant. When it comes to the null, $X>0$ since $C$ becomes larger than the capacitance that can make the reactance vanish. At this point, the current in a small coil is $-V_0/|X|$ and it has opposite direction to the current in the radiating loop antenna. Therefore, the reradiated magnetic field cancels the magnetic field radiated by the loop antenna. Similarly, when $X$ becomes even larger, the induced current in the small coil reduces dramatically and this alleviates the negative effect.

\begin{figure}[t]
  \centering
    \includegraphics[width=0.32\textwidth]{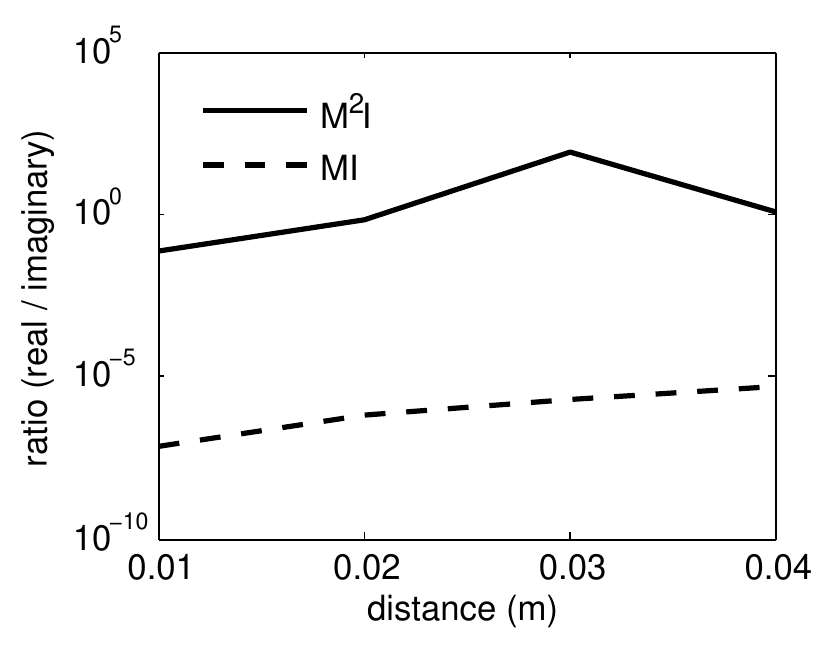}
    \vspace{-5pt}
  \caption{Real and imaginary power density change for MI and M$^2$I. The ratio is real power density over imaginary power density.}
  \vspace{-10pt}
  \label{fig:power_compare}
\end{figure}

In additional to the model-based understanding of the M$^2$I communication given in \cite{guo2015m2i}, here we can provide a more intuitive explanation on the enhancement mechanism of M$^2$I communication from the perspective of practical design and implementation. As discussed previously, the fundamental reason of the very limited communication range of the original MI is the very inefficient antenna, which is due to the extremely small coil size (denoted as $a$, 5~cm in this paper) compared with the signal wavelength (denoted as $\lambda$, tens of meters in HF band). Around the original inefficient MI antenna, the ratio of the reactive power to the real power is approximately $(\frac{\lambda}{2\pi a})^3$\cite{Ziolkowski_antenna_efficiency_2007}. Therefore, the majority of the power generated by the coil antenna is reactive, which only exists in the very close vicinity of the coil antenna and cannot be used for wireless communications if the transmission distance is relatively long.
In contrast, the M$^2$I communication can utilize such reactive power by converting a portion of the reactive power to the real power that the MI communication can rely on. Specifically, the reactive power induces current in the small coils on the metamaterial shell. The small coil then radiates real power and reactive power in the same fashion as the loop antenna. Due to the large number of small coils, the real power converted from the reactive power is significant compared with the original real power (three orders of magnitude in theory \cite{guo2015m2i} and one order of magnitude realized in this paper). This also explains why the metamaterial enhancement of M$^2$I apply to not only the near region but also the far region of the transmitting antenna. To validate the above discussion, in Fig.~\ref{fig:power_compare} we compare the ratio of real power density over imaginary power density of MI and M$^2$I when the distance is smaller than the metamaterial shell radius (inside the shell). Consistent with preceding discussions, the M$^2$I has much higher ratio that shows the imaginary power is converted into real power in the metamaterial shell. Note that, this enhancement will not change the antenna pattern but increases its radiation resistance. Through the simulation, we find that the radiation can be improved by around 10 times, which makes the antenna easier to match.

\subsection{Comparison with Ideal M$^2$I Communication}
In \cite{guo2015m2i}, we notice that the self-inductance can be negative around the system's optimal condition and in the complex and lossy medium the loop antenna's resistance can be increased dramatically. Now, with this realistic M$^2$I shell, we reinvestigate the above effects and show the fundamental physics behind them.
\subsubsection{Self-Inductance of M$^2$I Antenna}

\begin{figure}[t]
  \centering
  \subfigure[Capacitance is 2.5~nF]{
    \label{fig:fd0}
    \includegraphics[width=0.225\textwidth]{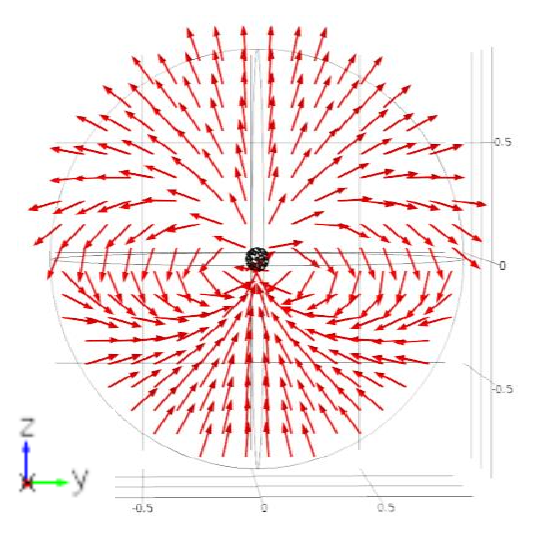}}\quad
  \subfigure[Capacitance is 2.55~nF]{%
    \includegraphics[width=0.225\textwidth]{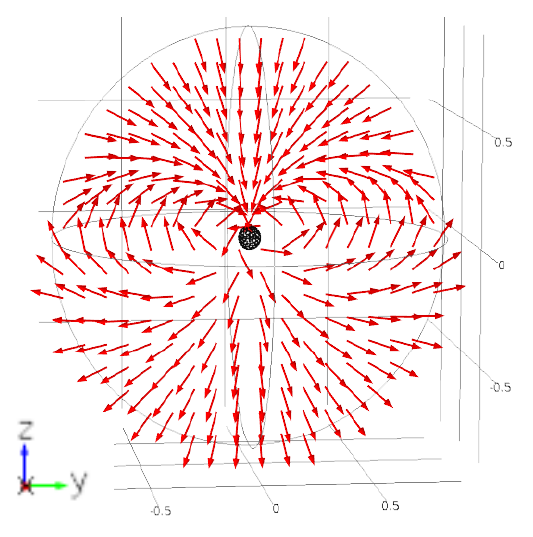}\quad
    \label{fig:fd1}}
    \subfigure[Capacitance is 2.6~nF]{%
    \includegraphics[width=0.225\textwidth]{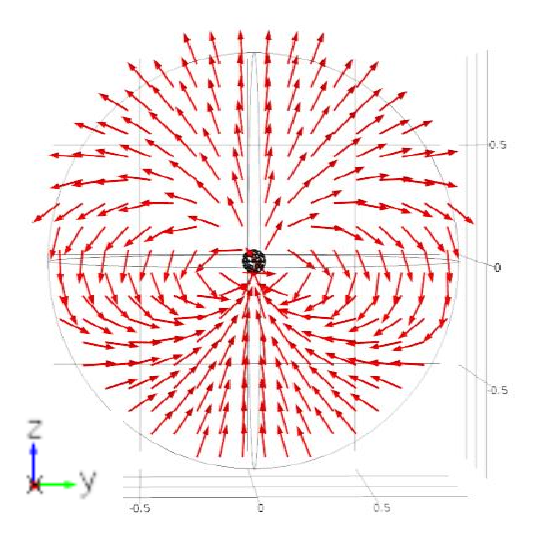}
    \label{fig:fd2}}
  \caption{Simulation results of magnetic field direction close to optimal condition.}
    \vspace{-15pt}
  \label{fig:fd}
\end{figure}
In \cite{guo2015m2i}, we found that the loop antenna's self-inductance can be negative and the reason is explained using EM theory. The antenna's self-inductance is expressed as $L_m=L_0+\Delta L$, where $L_m$ is the overall self-inductance, $L_0$ is the self-inductance without the metamaterial shell and $\Delta L$ is the additional self-inductance caused by the metamaterial shell. When the system is not resonant, $\Delta L$ is negligible and $L_m$ is positive. Near the optimal condition, $\Delta L$ changes dramatically. When $\Delta L$ is positive and has a large absolute value, $L_m$ is still a positive number, while when $\Delta L$ is negative and has a large absolute value, $L_m$ becomes negative and the loop antenna demonstrates a negative self-inductance. The results are also validated in Fig. 14 and Fig. 16 in \cite{guo2015m2i}. Here, by using the realistic M$^2$I, we analyze the negative self-inductance under practical conditions.

As analyzed in preceding subsection, the radiated magnetic field can be either enhanced or canceled by the metamaterial shell. Using 20~MHz as an example, we conduct the full-wave simulation and the results are shown in Fig. \ref{fig:fd}. The input current is positive and thus the magnetic field along the axis of the coil is positive z-direction. As shown in Fig.~\ref{fig:fd0}, when capacitance is 2.5~nF, the magnetic field outside the shell is the same direction as that without the shell, i.e., the field is enhanced by the shell. Referring back to Fig.~\ref{fig:radiation_20MHz}, this is around the peak. Effectively, it is equivalent to the condition that the M$^2$I loop antenna has better mutual coupling than MI and thus its self-inductance is increased. When the capacitance becomes 2.55~nF, the magnetic field outside the shell changes its direction by 180 degree because of that the reradiated field by the small coils are larger than the original field radiated by the loop antenna. Therefore, the loop antenna radiates magnetic field, which obeys left-hand rule and its self-inductance is negative. When the capacitance is 2.6~nF, the coils on the shell is far from resonance and, as a result, its reradiated field becomes weak and the field radiated by the loop antenna is dominant again. Thus, the magnetic field direction is the same as that without the shell.

\begin{figure}[t]
  \centering
    \includegraphics[width=0.3\textwidth]{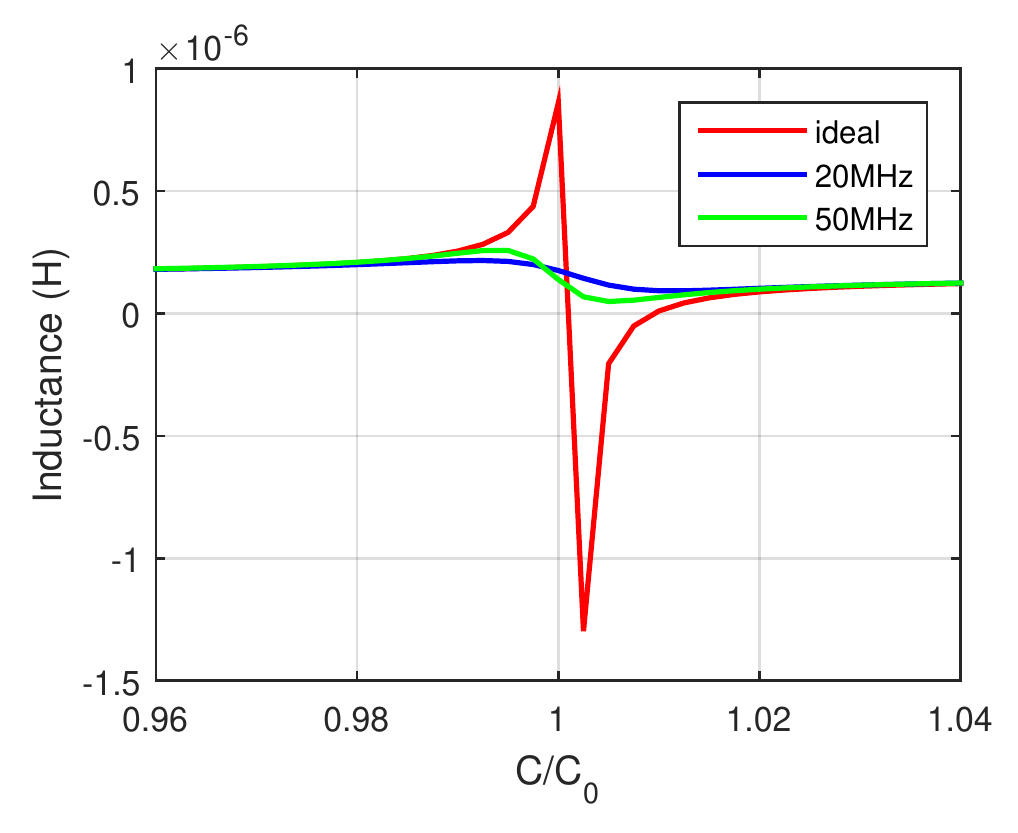}
    \vspace{-15pt}
  \caption{Effect of unit coil's capacitance on loop antenna's self-inductance }
  \vspace{-10pt}
  \label{fig:inductance}
\end{figure}

Besides 20~MHz with $C_0$=2.16~nF and 50~MHz with $C_0$=0.353~nF, we consider an ideal scenario where the frequency is 20~MHz, $C_0$ is 2.16~nF, but the coil resistance $R_c$ is reduced 1000 times which is almost equivalent to a perfect no-loss metamaterial. The effective self-inductance can be expressed as $L_{eff}=L_0+\Im({Z_{ref}})/j\omega$, where $\Im$ denotes the imaginary part of a complex number. $L_{eff}$ is also plotted in Fig. \ref{fig:inductance}. Due to the coupling among the loop antenna in the shell and coils on the shell, the reflected impedance in the loop antenna need to be considered, which can be expressed as~\cite{Sun_MI_TAP_2010} $Z_{ref}=\sum_{i=1}^{N} {\omega^2 M_{0i}^2}/{Z_i}$, where the subscript 0 denotes the loop antenna. The negative self-inductance generated by the spherical coil array is consistent with the discussion in \cite{guo2015m2i}, i.e., Fig. 14 and Fig. 16, by using the ideal metamaterial. Notice that, the ideal scenario can achieve a negative self-inductance as predicted. However, 20~MHz and 50~MHz can only achieve relatively small self-inductance, but cannot be negative. The reason is that their resonance is not as strong as the ideal scenario due to the coil resistance. Inside the shell, the dominant field is from the loop antenna since the reradiated field from the small coils are not strong enough, while outside the shell the field radiated by the small coils are larger than the loop antenna since the small coils are closer to the observation point. As a result, although $\Delta L$ becomes negative, the reflected impedance in the loop antenna is not large enough to cancel the antenna's positive self-inductance.

\subsubsection{M$^2$I Communication in Lossy Medium}
\label{sec:lossy}
The loss of the metamaterial shell is due to the resistance of the small coil on the shell. In previous discussions, the environment is considered as lossless, while the original MI communication is proposed to improve the communication performance in lossy underground environment. If M$^2$I works in a lossy medium, the performance would be affected by the environment's conductivity, because of two reasons. First, due to the lossy medium, in the near region of the small coil on the shell, there is an additional loss which can be regraded as a reflected impedance in the coil. As a result, the overall resistance of the coil increases. This effect is more significant when the conductivity of the medium is high. The large resistance of the coil results in a high-loss metamaterial effectively, which can dramatically reduce the communication range or channel capacity. Note that, in \cite{guo2015m2i} the ideal metamaterial is assumed to have a constant loss which is not affected by the medium's conductivity. However, for realistic M$^2$I this does not hold since the metamaterial elements are also affected by the medium's loss. Second, the lossy medium can affect the radiating loop antenna in the same manner as affecting the small coils, i.e., the loop antenna also has an additional resistance caused by the medium, which induces more loss in the antenna.

This challenge can be well addressed by using a larger spherical shell to enclosed the M$^2$I antenna. As shown in \cite{tai1947hertzian,wait1952magnetic,james1957insulated}, the additional resistance is the reciprocal of the radius of the dielectric sphere. In other words, the larger the sphere we used to enclose the M$^2$I antenna, the smaller the additional resistance. Here, we just provide this solution, the detailed modeling is out of the scope of this paper since it involves spherical wave-based EM field analyses.

\begin{figure}[t]
  \centering
    \includegraphics[width=0.3\textwidth]{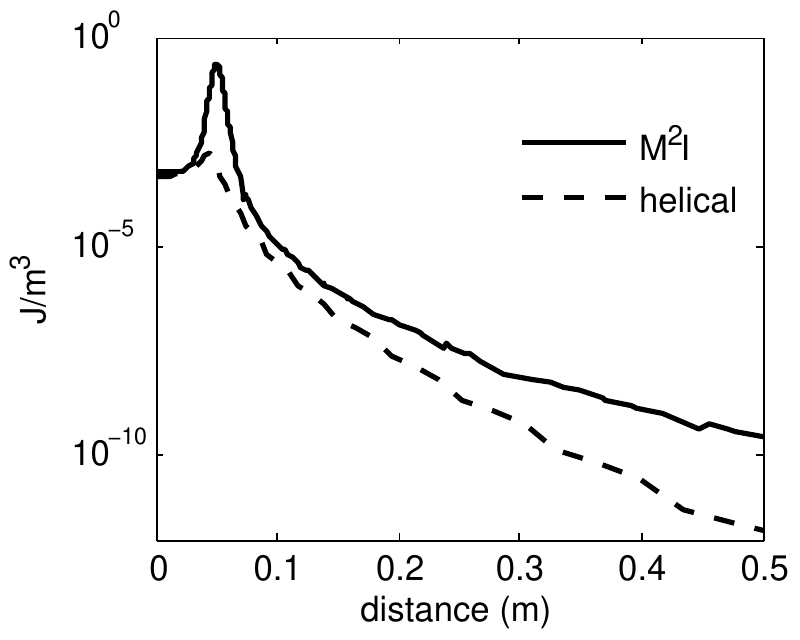}
    \vspace{-10pt}
  \caption{Generated energy density of M$^2$I and spherical helix antenna.}
  \vspace{-7pt}
  \label{fig:helix}
\end{figure}

In addition, the reason we employ magnetic dipole antenna rather than other typical electric antennas is due to its attracting properties in lossy medium, such as stable channel, low multipath fading, and low absorption loss \cite{Sun_MI_TAP_2010,gulbahar2012communication}. According to the theoretical results in \cite{karlsson2004physical}, magnetic dipole is the most efficient antenna in lossy medium, especially when the antenna size is extremely electrically small. In lossy media, such as underground and underwater, the magnetic dipole has been extensively used \cite{gulbahar2012communication,sun2011dynamic,kisseleff2015digital,large1973radio}. In Fig.~\ref{fig:helix} we provide the numerical simulation results to compare the energy density generated by M$^2$I with a spherical helix antenna. The antenna is configured the same as the 7-turn antenna in \cite{hui2001input}, but its radius is changed to 5~cm to make it the same as the M$^2$I. Since the spherical helix antenna with 5~cm radius resonant at much higher frequency (308 MHz) than M$^2$I, the energy density attenuates very fast. The M$^2$I using magnetic induction experience less absorption.
\subsection{Path Loss and Channel Capacity}
Up to this point, we have demonstrated that the coil array-based M$^2$I can significantly enhance the radiated magnetic field intensity. In this section, we present the performance of the wireless channel between two M$^2$I transceivers, i.e., both the transmitter and receiver are equipped with coil array-based M$^2$I antenna.
In this research, the path loss and channel capacity are employed to evaluate the performance of M$^2$I. In the terrestrial antenna design, the radiation efficiency is an important parameter since it reflects the capability of the antenna to transmit power into the far field\cite{Balanis_a}. However, in MI communication, both the reactive power and near region real power are utilized and thus the radiation efficiency can not comprehensively capture the communication capability of the antenna. The path loss is a widely used parameter in MI communication because it considers the received power including both reactive and real power \cite{Sun_MI_TAP_2010}.

\begin{figure}[t]
  \centering
    \includegraphics[width=0.45\textwidth]{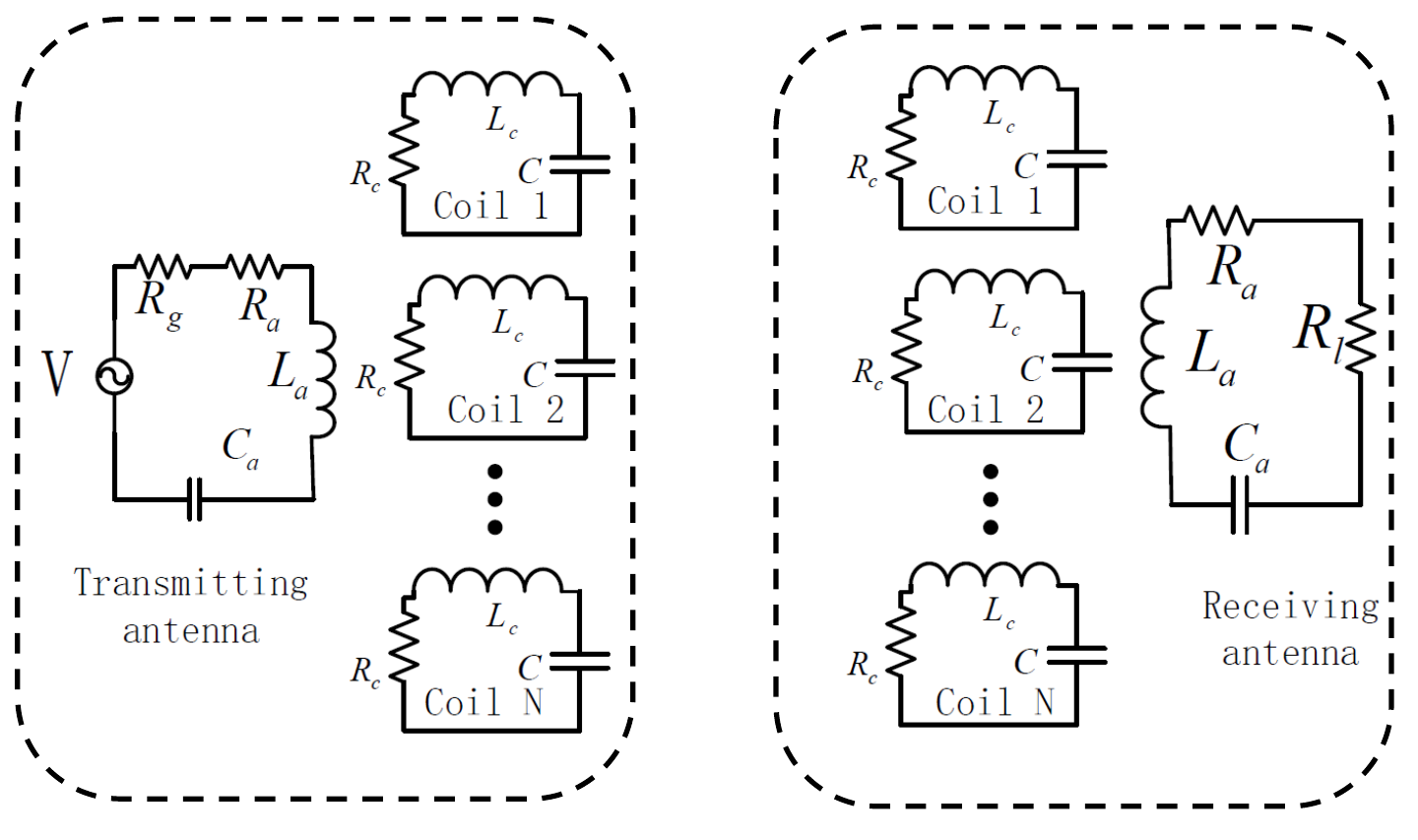}
    \vspace{-10pt}
  \caption{ Equivalent circuit: transmitting antenna and receiving antenna with coils on their M$^2$I shell. }
  \vspace{-12pt}
  \label{fig:circuit}
\end{figure}

\begin{figure}[t]
  \centering
    \includegraphics[width=0.4\textwidth]{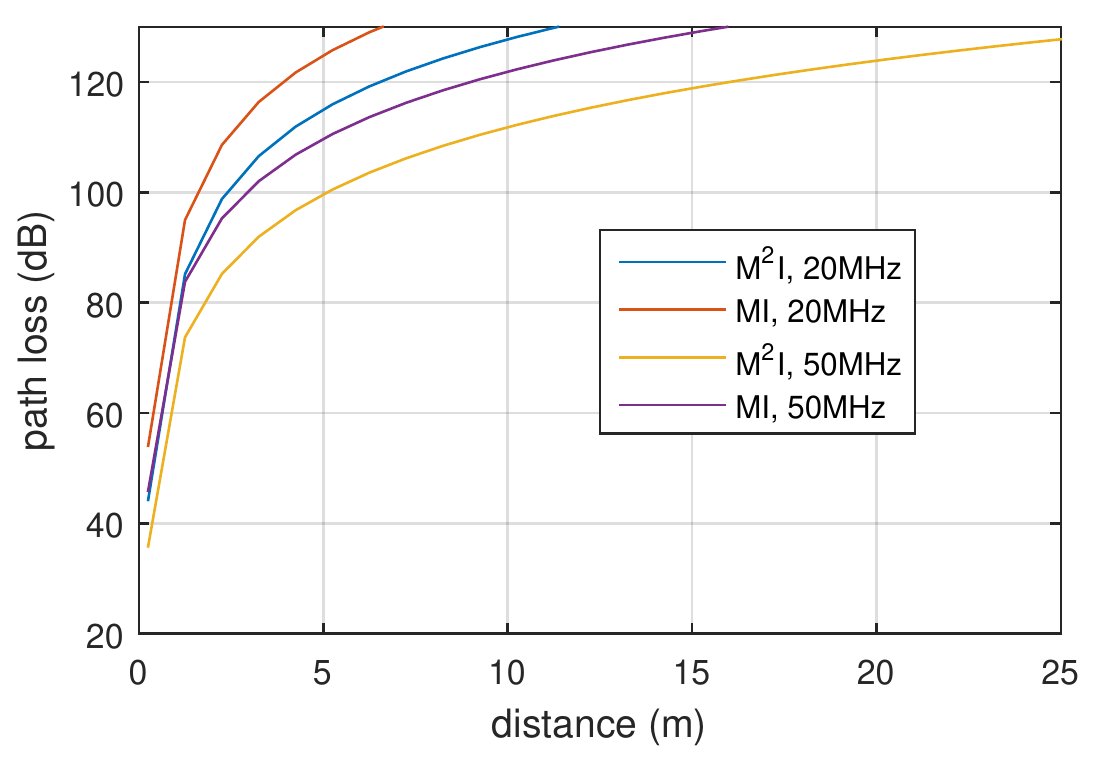}
    \vspace{-10pt}
  \caption{Path loss of MI and M$^2$I. Distance is from transmitter towards receiver.}
  \vspace{-13pt}
  \label{fig:pathloss}
\end{figure}

\begin{figure}[t]
  \centering
  \subfigure[20~MHz]{
    \label{fig:return_loss_20MHz}
    \includegraphics[width=0.22\textwidth]{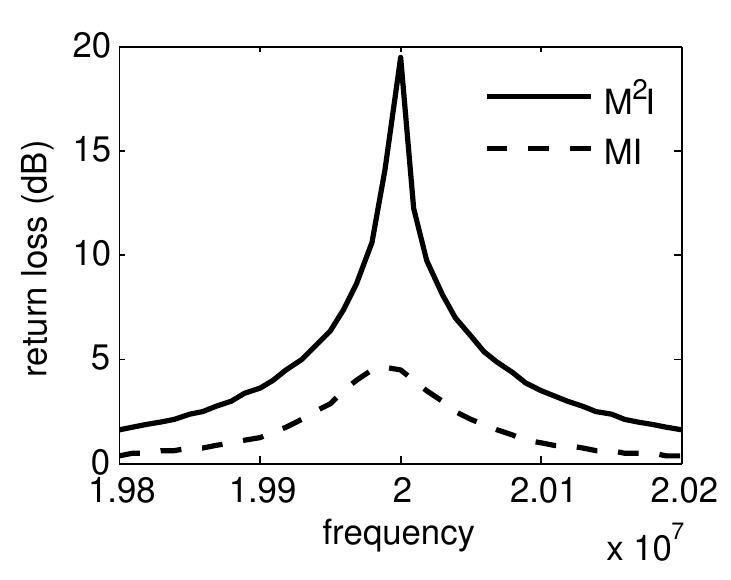}}\quad
  \subfigure[50~MHz]{%
    \includegraphics[width=0.22\textwidth]{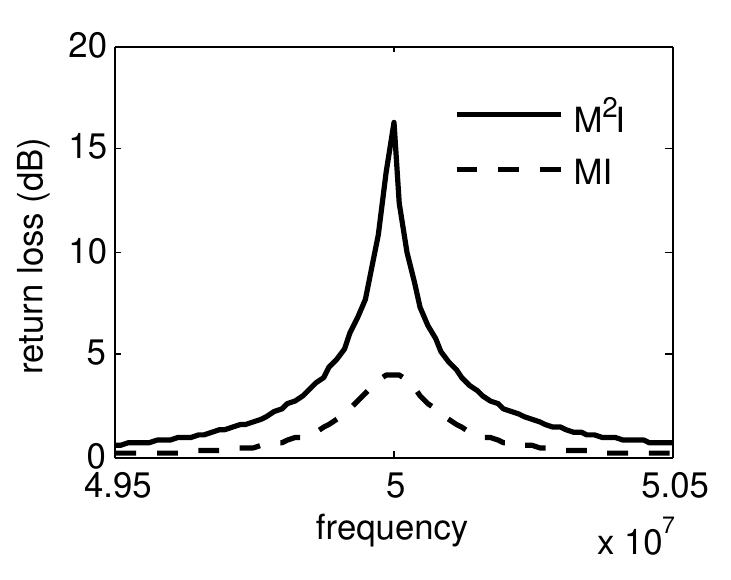}\quad
    \label{fig:return_loss_50MHz}}
  \caption{Return loss of MI and M$^2$I. }
    \vspace{-15pt}
  \label{fig:return_loss}
\end{figure}

\begin{figure}[t]
  \centering
    \includegraphics[width=0.4\textwidth]{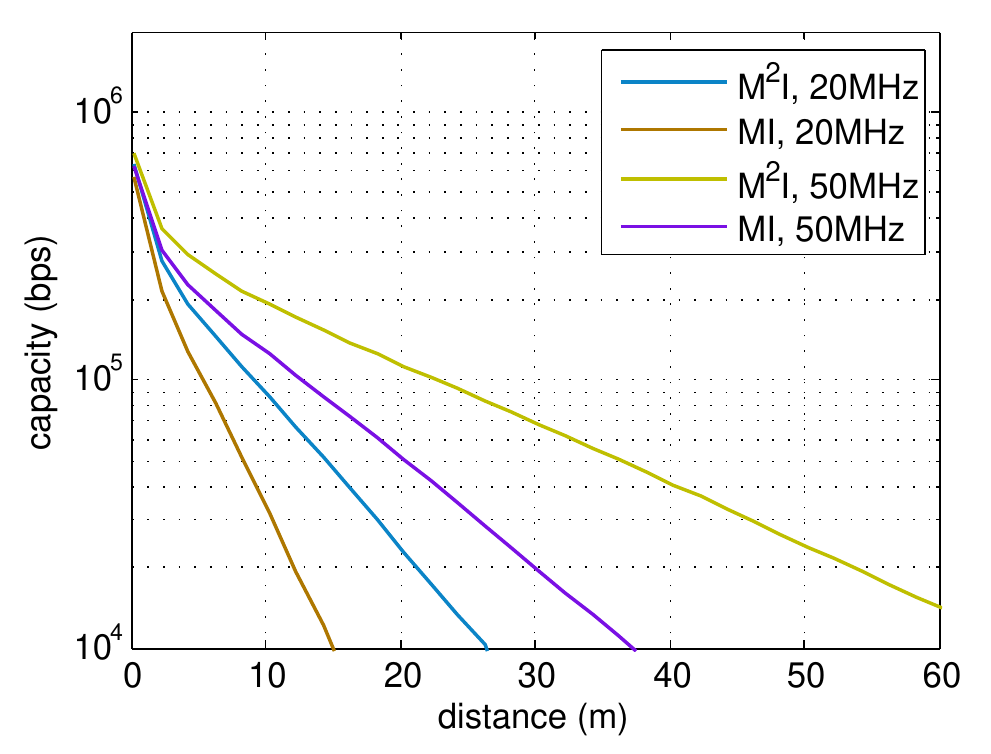}
    \vspace{-10pt}
  \caption{Channel capacity for M$^2$I and MI. Distance is from transmitter towards receiver.}
  \vspace{-13pt}
  \label{fig:capacity}
\end{figure}
The equivalent circuit is shown in Fig. \ref{fig:circuit}, where $R_g$ and $R_l$ are the source's output resistance and load resistance, respectively. Both of the $R_g$ and $R_l$ are set as a typical value 50~$\Omega$. The resistance, self-inductance, and tunable capacitance in the antenna circuit are denoted by $R_a$, $L_a$, and $C_a$, respectively.  As shown in the figure, the small coils on the shell are modeled as simple RLC circuit surround the loop antenna. When we compare the M$^2$I transceivers and original MI transceivers, the same as previous discussions, the radius of the original MI antenna is the outer radius of the shell, i.e., 0.05~m, while the M$^2$I antenna has a radius of 0.04~m and its shell radius is 0.05~m. In order to calculate the magnetic field at the receiving side, we need to add the passive coils in the receiving M$^2$I antenna into \eqref{equ:app1} and \eqref{equ:field}.
%Now, we have a transmitting antenna, 84 passive coils, and a receiving antenna. Therefore, the updated dimensions of ${\bf Z}$, ${\bf I}$ and ${\bf V}$ in \eqref{equ:app1} are 86$\times$86, 86$\times$1, and 86$\times$1, respectively.

The voltage for the transmitting loop antenna is $V_0$. All other voltages in the updated \eqref{equ:app1} are 0. By solving the updated \eqref{equ:app1}, the induced current in antennas can be found. The received power can be expressed as $P_r=1/2|I_l|^2R_l$, where $I_l$ is the load current. Similarly, the dissipated power in the source can be expressed as $P_t=1/2\Re{(I_t V_0)}$. Then, we can obtain the path loss ${\mathcal L}(d)=-10\log(P_r/ P_t)$. Next, by using the optimal capacitance value 2.16~nF for 20~MHz and 0.353~nF for 50~MHz, we compare the path loss of M$^2$I and MI. The loop antenna's resistance and self-inductance can be found in \cite[Ch.5]{Balanis_a}. As shown in Fig. \ref{fig:pathloss}, the path loss of M$^2$I is much lower than that of original MI which is consistent with~\cite{guo2015m2i}. Also, the 50~MHz achieves lower path loss than 20~MHz since the antennas have stronger coupling at higher frequency. Here, the considered medium is lossless, while when the medium becomes lossy, higher frequency may not be a good choice since it suffers from higher absorption rate.

The return loss bandwidth is provided in Fig.~\ref{fig:return_loss}. As shown in the figure, the M$^2$I has much broader bandwidth than MI. The reason is that the MI antenna is electrically small and it has very small resistance, which is hard to be matched. Moreover, due to the small resistance, high quality factor impedance matching network is needed, which in turn reduces its bandwidth. On the contrary, the M$^2$I has relatively larger resistance due to the coupling between the loop antenna and the small coils on the shell. This makes the antenna easier to be matched. Also, the antenna has broader bandwidth than the MI. The bandwidth exceeds the physical limit is because of the following reasons. First, since the electrical size is extremely small, the bandwidth is supposed to be very narrow. However, such narrow bandwidth is not suitable for wireless communication. Therefore, we need to balance the antenna efficiency and bandwidth. To maintain a reasonable bandwidth, we have to either add additional resistance or match the antenna imperfectly.

The Shannon channel capacity is also evaluated. According to \cite{rappaport1996wireless}, ${\mathcal C}=Bw \cdot \log_2[1+P_r/(Bw\cdot N_{noise})]$, where $Bw$ is the bandwidth and $N_{noise}$ is the noise density. Since in complex environments communication the bandwidth is usually smaller than 20~kHz \cite{pittman1985through} and both MI and M$^2$I can meet this requirement, the bandwidth is set as 20~kHz. Also, the transmission power is 10~dBm. Besides the thermal noise, additional metamaterial noises can be introduced by the small coils on the shell and they are treated by following the method in \cite{wiltshire2014noise,stevens2010magnetic}. As shown in Fig. \ref{fig:capacity}, M$^2$I can increase the communication range significantly. Note that in the near field the capacity is constrained by bandwidth since the received power is strong enough and thus the performances are similar. When the distance becomes large, the capacity is constrained by the received power since the signal to noise ratio is much smaller than that in the near field. As a result, the gain of M$^2$I becomes large.

\section{Implementation and Experimental Analysis}
In this section, we implement the M$^2$I antenna using a 3D printed spherical frame and printed coils on circuit boards. The magnetic field enhancement is validated and the wireless channel between M$^2$I transceivers are measured in different environments.
\begin{figure}[t]
  \centering
  \subfigure[]{
    \label{fig:3d_printing_1}
    \includegraphics[width=0.18\textwidth]{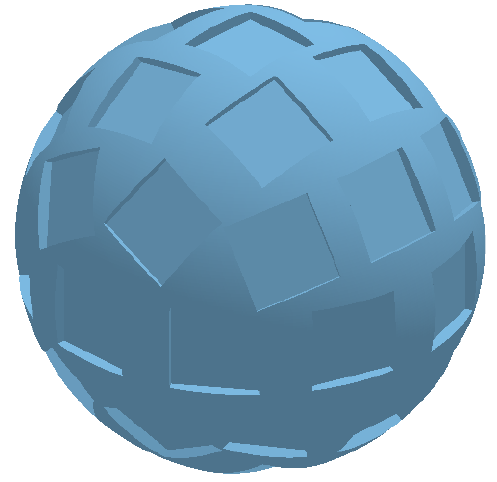}}\quad\quad
  \subfigure[]{
    \label{fig:3d_printing_2}
    \includegraphics[width=0.18\textwidth]{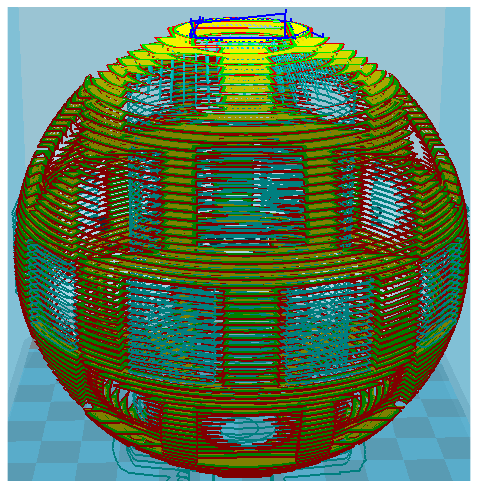}}\quad\quad
  \subfigure[]{%
    \includegraphics[width=0.32\textwidth]{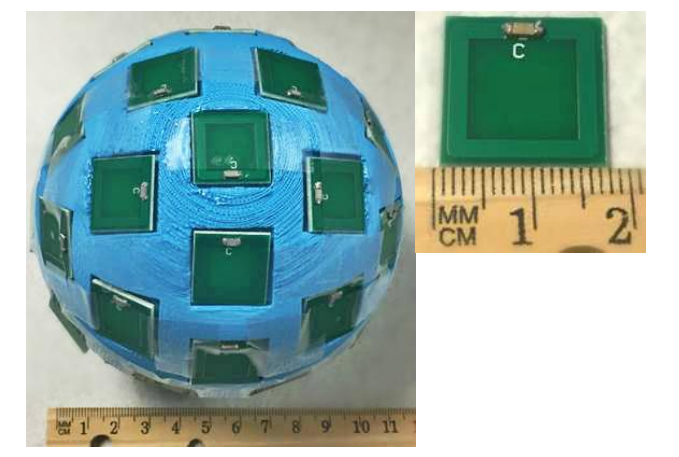}\quad\quad
    \label{fig:sphere_exp}}
      \vspace{-5pt}
  \caption{M$^2$I prototype with 3D printed sphere and coil antennas. (a) 3D digital model; (b) sliced model; (c) implementation}
    \vspace{-15pt}
  \label{fig:3D_printing_implementation}
\end{figure}
\subsection{Antenna Implementation for M$^2$I Communication}
%The M$^2$I antenna consists of the 3D printed spherical frame and the metamaterial units.

\subsubsection{3D Printed Spherical Frame}

A spherical frame is fabricated by using 3D printing technique to support the coil-antenna array. 3D printing is an emerging disruptive technology that can directly fabricate 3D object from digital model without specific tooling and fixturing. Therefore it is more flexible and efficient than traditional molding and machining based manufacturing processes. Fused Deposition Molding (FDM) is one of most popular 3D printing technology which is advantageous in terms of material property, cost and accessibility compared with other technologies. In this research, FDM based desktop printer (MakerBot Replicator 2X) has been employed to fabricate the spherical shape support structure for the coil-antenna array.
This process begins with a 3D digital model representing the geometry and topology of the final product, which is modeled in a commercial Computer-Aided Design (CAD) software suite (Creo Pro, PTC) as shown in Fig.~\ref{fig:3d_printing_1}. Dedicated model pre-processing software (Makerware) slices the digital model into a series of 2D layers with several micro-meter thicknesses, and then rasterizes the layers by tool path planning (Fig.~\ref{fig:3d_printing_2}). The path data is then transmitted to the 3D printer and the system operates in X, Y and Z axes, drawing the model one layer at a time. The material is fed into the printer in the form of filament, and then heated up to the material melting point and extruded through nozzle with small orifice. The spherical frame is fabricated in a layer by layer basis.

\subsubsection{Metamaterial Units and Coil-antenna Array}
As discussed in the previous sections, the spherical array of small coils is the effective component to realize the metamaterial layer for M$^2$I antenna. In our implementation, we place the small coils on the 3D printed frame, which forms the spherical coil-array. The spherical frame and the small coils have exactly the same size as the simulation model in previous section. The small coils are printed on circuit boards with thickness 1.57~mm. The thickness of the copper is 1~oz. The edge of the square PCB is 18~mm long. The outer edge of the coil is 16~mm and the width of the trace is 2~mm. All the coils are placed at the same position as the simulation model. The final product of the 3D printed spherical frame with metamaterial units (i.e., the small coils) is shown in Fig.~\ref{fig:sphere_exp}.

\subsection{Wireless Coupling Enhancement}
As discussed in Section~\ref{sec:rfa}, the radiated magnetic field by a magnetic dipole can be enhanced by the metamaterial shell. To validate this conclusion, we resort to the $S_{21}$ parameter \cite{pozar2009microwave} since it is proportional to the ratio of transmitted voltage to induced voltage; as indicated in \eqref{equ:voltage}, the voltage is proportional to magnetic field and thus the $S_{21}$ is a good indicator to show the enhancement of wireless coupling. Analogous to the method used in \cite{scarborough2012experimental,lipworth2015quasi} to evaluate metamaterial's performance, we use a pair of nonresonant magnetic loop antennas with radius 0.045~m and measure the $S_{21}$ parameter with metamaterial sphere ($S_{21}^w$) and without metamaterial sphere ($S_{21}^{wo}$). We define the $S_{21}$ gain as $S_{21}^w-S_{21}^{wo}$ where the $S_{21}$ parameters are in dB scale. The $S_{21}$ parameter is measured in Agilent 8753E RF network analyzer. We use 2700~pF capacitors to tune the small coils on the sphere.

\begin{figure}[t]
  \centering
    \includegraphics[width=0.36\textwidth]{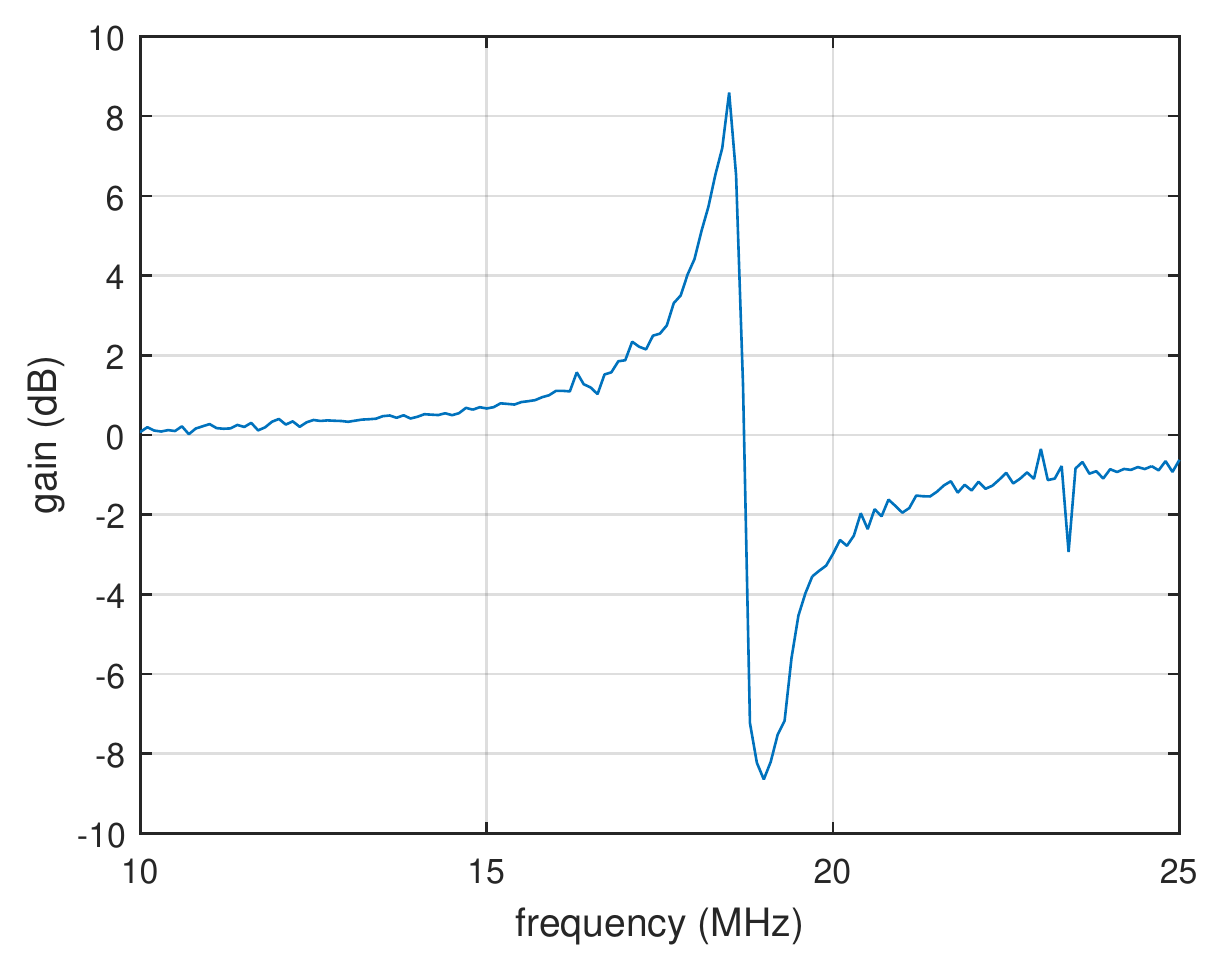}
    \vspace{-10pt}
  \caption{$S_{21}$ parameter gain. The $S_{21}$ of M$^2$I in dB scale minus the $S_{21}$ of MI in dB scale. }
  \vspace{-15pt}
  \label{fig:S21_gain}
\end{figure}

As shown in Fig.~\ref{fig:S21_gain}, a resonance is achieved at 18~MHz and the gain is around 8~dB. The results show that both the resonant peak at 18~MHz and the null right after it are exactly the same as our analyses and prediction. It is worth noting that the implementation of M$^2$I requires the small coils on the shell to be highly identical, i.e., the capacitors should have low tolerance. The capacitors we use have 1\% tolerance. If the tolerance is higher, the small coils on the sphere have slightly different resonance frequency and thus the negative permeability on the sphere are inhomogeneous, or even worse some of the coils may demonstrate positive effective parameter. Similarly, the coils are supposed to be identical, otherwise their self-inductance and resistance are different, which can also create inhomogeneity on the sphere.

\subsection{Wireless Channel Measurement}
To measure the path loss the of wireless channel between two M$^2$I transceivers, we use USRP software-defined radio kits as the signal transmitting and receiving devices. The M$^2$I antennas are connected to the USRP boards. The mother board we utilized is the USRP N210, which is based on a Xilinx Spartan-3A DSP 3400 FPGA. It has a 100~MS/s dual ADC and a 400~MS/s dual DAC, and it is connected to computer via a gigabit ethernet. The daughter boards are LFTX/LFRX which can support two independent antennas through connectors TxA/RxA and TxB/RxB. This daughter board can generate and receive wireless signals from 0~MHz to 30~MHz which is within our interest frequency band, i.e., 18~MHz. Based on these hardware equipments, the signal is generated and analyzed in GNU softwares installed on a computer. The signal is generated by a signal source block with frequency 100~kHz and the transmitting frequency is set as 18~MHz. At the receiving side, we use a fast Fourier transformation (FFT) block to convert the received time domain signal into frequency domain signal. Note that, the received power here is in dB scale rather than dBm since it is a relative value which is determined by the device. The exact power value in dBm needs a comprehensive analysis on the power loss and gain of each component that the signal goes through. In the following, we provide the same transmitting power and measure the received power to show the difference between the case using M$^2$I antennas and the case using original MI antennas. Since the metamaterial shell is symmetrical, the shell have the same influence on the inner loop antenna with arbitrary orientation. In the experiments, the transmitting and receiving antennas are placed coplanarly. The experiments are conducted in two environments for the aforementioned applications, including indoor environment and underground environment.

\begin{figure}
  \centering
  \subfigure[Indoor]{%
    \includegraphics[width=0.42\textwidth]{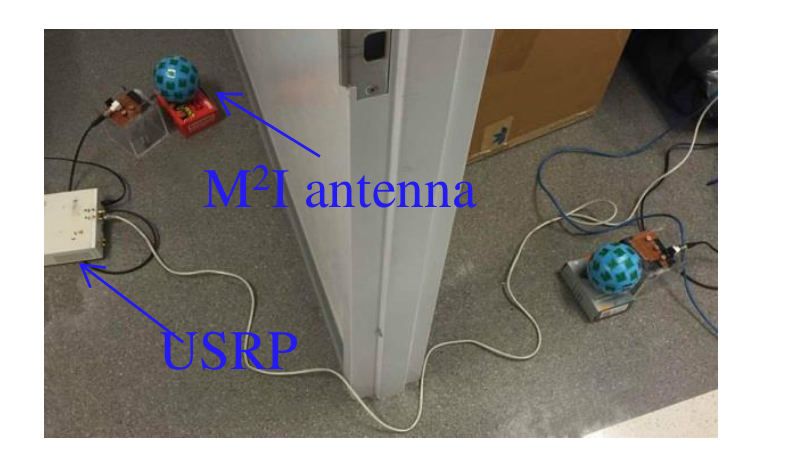}
      \vspace{-1pt}
    \label{fig:indoor}}
  \subfigure[Underground]{
    \label{fig:underground}
    \includegraphics[width=0.5\textwidth]{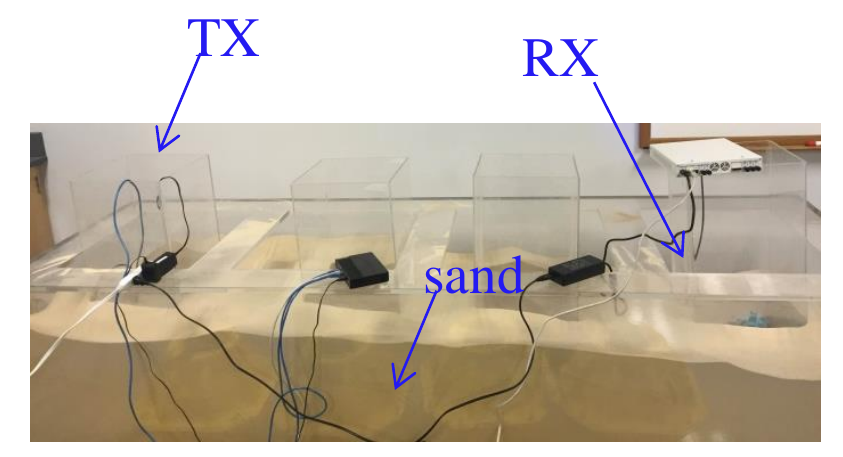}}
      %\vspace{-5pt}
  \caption{Experiment environments and setup. }
    \vspace{-8pt}
  \label{fig:communication_test}
\end{figure}

\subsubsection{Indoor}
\begin{figure}
\centering
    \includegraphics[width=0.4\textwidth]{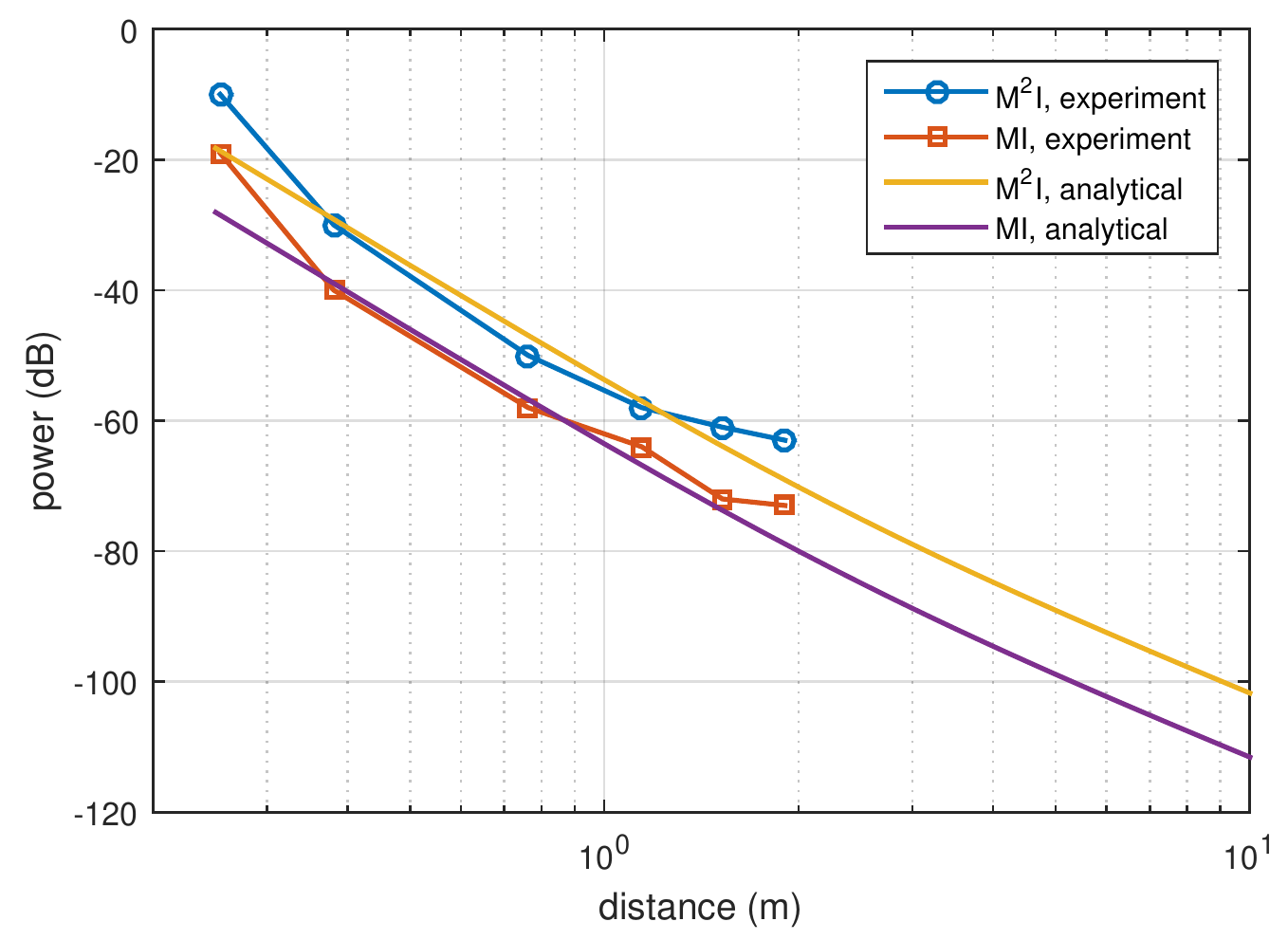}
    \vspace{-12pt}
  \caption{Received power in indoor environment. }
  \label{fig:indoor_power}
\end{figure}

\begin{figure}[t]
  \centering
  \subfigure[With metamaterial]{
    \label{fig:15in_wo_mm}
    \includegraphics[width=0.32\textwidth]{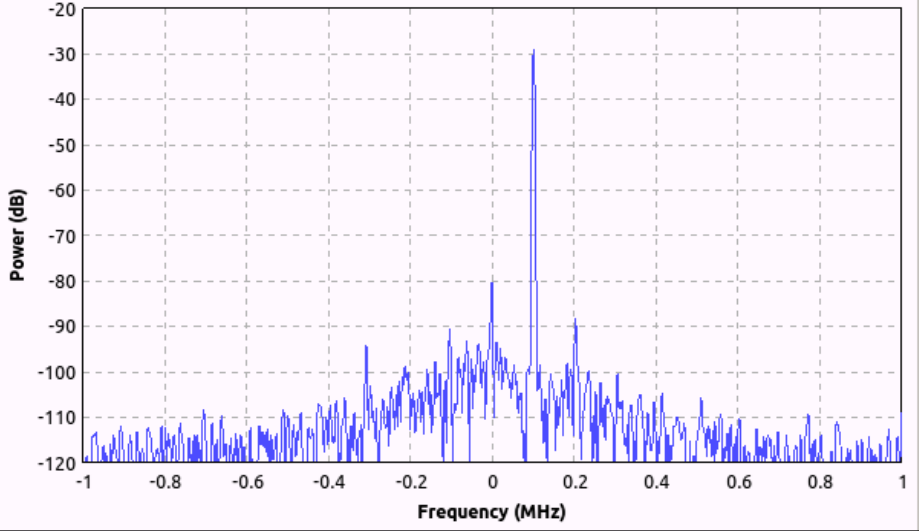}}\quad
  \subfigure[Without metamaterial]{
    \label{fig:15in_wi_mm}
    \includegraphics[width=0.32\textwidth]{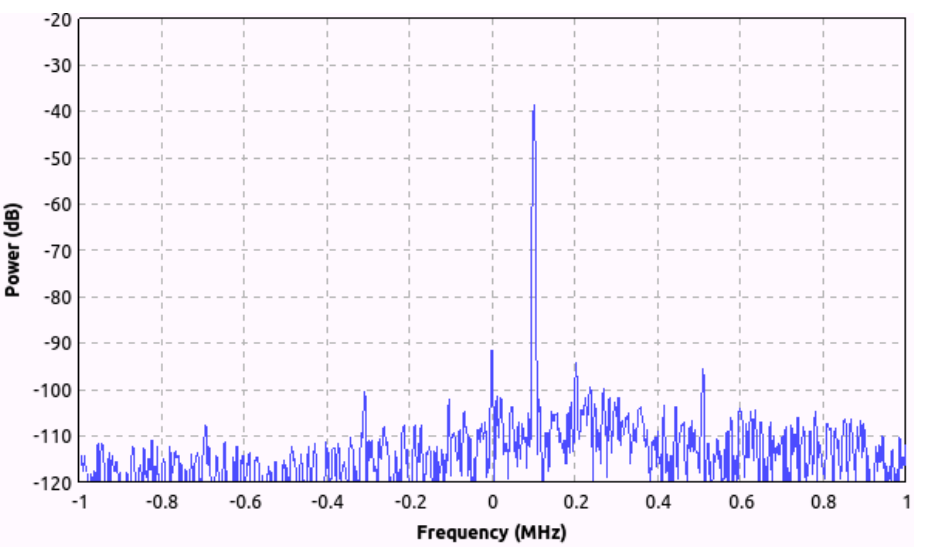}}\quad
      \vspace{-2pt}
  \caption{Measured received power at 15~in from transmitter. }
    \vspace{-10pt}
  \label{fig:communication}
\end{figure}

We first test the path loss in indoor environments and the settings employed is depicted in Fig.~\ref{fig:indoor}. By varying the distance between a transmitter and a receiver, we measure the received power of using metamaterial sphere and without it. As shown in Fig.~\ref{fig:indoor_power}, with the metamaterial shell, M$^2$I can receive around 10~dB more power than original MI and the communication range can be extended. Also, as an example, the measured power at 38~cm (15~in) is shown in Fig.~\ref{fig:communication} for both M$^2$I and MI. The power spectrum clearly shows that the baseband signal strength is increased by around 10~dB (approximately one order of magnitude) by using metamaterial sphere. In another word, if we use the same transmission power and the same receiver sensitivity, the 10 dB enhancement of M$^2$I can almost double the transmission range of the original MI. In Fig.~\ref{fig:indoor_power}, the analytical path losses predicted by the theoretical model in \cite{guo2015m2i} are also depicted for comparison, which has a good match with the experiment results.

\subsubsection{Underground}
\begin{figure}
\centering
  \includegraphics[width=0.25\textwidth]{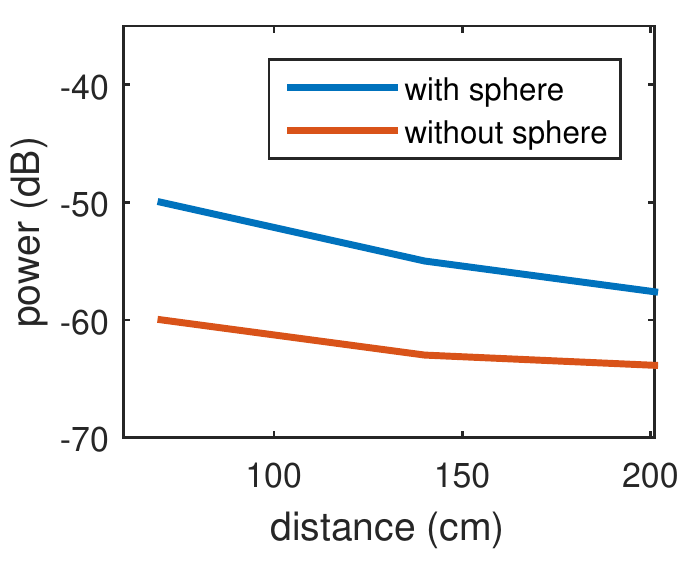}
  \vspace{-7pt}
  \caption{Received power in underground environment.}
    \label{fig:ug}
    \vspace{-10pt}
\end{figure}

To test M$^2$I's performance in complex environments, a tank with length 255~cm, height 76~cm and width 76~cm is utilized and filled with around 0.98~m$^3$ of sand. The USRP and M$^2$I antennas are placed in the sand to test its performance in the underground environment. We measure the received power at 70~cm, 140~cm, and 200~cm away from the transmitting antenna. As shown in Fig.~\ref{fig:ug}, with the metamaterial shell, the received power are -50~dB, -55~dB, and -58~dB, while without the shell, the received power are -60~dB, -63~dB, and -64~dB. Hence, the gain is also significant in such environment. It should be noted that the soil tank actually creates a confined space with six soil-air boundaries, which is much more complicated than the homogeneous environment used in the theoretical model in \cite{guo2015m2i}. As a result, the theoretical curves are not compared here.

%Since sand has higher dielectric parameter and it is dry which has almost no loss, the field is a little stronger than that in air.

%\subsubsection{Wall}
%One of the important applications of MI communication is the indoor localization \cite{tan2015environment}. However, the walls in indoor environment is a big enemy for wireless signals. Next, we put the M$^2$I transmitter and receiver in different rooms as shown in Fig.~\ref{fig:wall} to show its advantage over MI. The door is closed when we measure the power. By using the same devices and configurations as previous discussions, with the shell the received power is -50~dB, while without the shell the power is -55~dB. Through the experiment, it shows that although the received power gain is still large but it becomes smaller than that in indoor environment. The reason is that the wall and the door have a certain of loss, which can enlarge of the loss of the metamaterial shell by increasing the coils' reflected impedance and thus reduce the performance of M$^2$I, as discussed in previous section.

\section{Conclusion}
Magnetic Induction (MI) communication can enable a large number of novel applications in complex environments, which can significantly enrich the wireless connectivity. Although it has been proved to be a promising solution, its communication capability is still limited by its antenna efficiency. Recently, metamaterial has been introduced to enlarge MI communication range and data rate. However, the existing works are based on ideal metamaterials which do not exist in nature. In this paper, we propose a practical design of Metamaterial-enhanced Magnetic Induction (M$^2$I) communication by using a spherical coil array. The physical principles and geometric structure of this design are introduced and its optimal configurations are found. The relation between this practical M$^2$I and the ideal metamaterial based M$^2$I are discussed. Through the communication performance evaluation, we find that the realized M$^2$I can significantly increase the channel capacity and thus extends the communication range, which validates the results promised by ideal M$^2$I. The proposed M$^2$I communication is also implemented and tested in various environments.

It should be noted that, compared with the theoretical prediction, i.e., three orders of magnitude enhancement, in \cite{guo2015m2i}, there still exists a performance gap in the practical design and implementation in this paper (one order of magnitude). We expect the performance of coil array based M$^2$I can be further improved through the following ways. First, the single layer shell can be extended to multiple layers to involve more metamaterial elements. Second, the isotropic metamaterial can be utilized to enhance the polar direction magnetic field radiated by the loop antenna since currently only the radial direction is enhanced which underestimates the loop antenna's performance. Third, current implementation still suffers from high loss. This can be solved by either reduce the resistance of the PCB coil or increase its turns to enlarge its quality factor. Moreover, active circuit can be leveraged to dramatically mitigate this effect.

%This design works for radial direction since the orientation of the coils on the shell is only radial direction. In order to make this enhancement isotropic, in our future work, a tri-directional coil will be added and experiments will be conducted to evaluate its performance. Then, the gap between ideal metamaterial and this spherical coil array can be closed.

\section*{Appendix}
\subsection{Mutual Inductance}
The mutual inductance utilized in \cref{equ:rearrange1,equ:effmu,equ:app1} is derived here. The mutual inductance between coil 0 and coil 1 can be denoted as $M_{01}$, where coil 0 is transmitting coil and coil 1 is receiving coil. The magnetic field radiated by coil 0 can be written as \cite[Ch.5]{Balanis_a},
\begin{align}
\label{equ:antenna_field}
H_r(I_0){\hat r_0}&=\frac{j k a^2 I_0 \cos{\theta}}{2d^2}\left[1+\frac{1}{jkd}\right]e^{-jkd}{\hat r_0},\\
H_{\theta}(I_0){\hat \theta_0}&=\frac{- k^2 a^2 I_0 \sin{\theta}}{4d}\left[1+\frac{1}{jkd}-\frac{1}{(kd)^2}\right]e^{-jkd}{\hat \theta_0},
\end{align}
Let the magnetic field intensity at coil 1 be $H_0 {\hat h}=H_r(I_0){\hat r_0}+H_{\theta}(I_0){\hat \theta_0}$, where ${\hat h}$ is the magnetic field direction. Then the mutual inductance can be expressed as $M_{01}={\mu_0 H_0 \pi a_1^2 {\hat h}\cdot {\hat o_1}}/{I_0},$
where $a_1$ is the radius of coil 1, $I_0$ is the current in coil 0 and ${\hat o_1}$ is the orientation vector of coil 1.
\subsection{Proof of Proposition \ref{pro:IH}}
The equation \eqref{equ:app1} can be written as ${\bf Z}{\bf I}_0={\bf V}_0$. We first consider ideal isotropic radiation source and then the analysis is extended to real dipole source. By assuming that all the elements in ${\bf V}_0$ are the same, we have $V_1=V_2=\cdots=V_N=-j\omega\pi a^2 \mu_0 {\hat H_r}$. Since all the coils are uniformly distributed on the spherical shell, the structure is symmetrical. Moreover, as the coils have the same excitation voltages, they should have the same induced currents, i.e., $I_1=I_2=\cdots=I_N$. Then, we have $I_1(Z_1+\sum_{i\neq 1}^{N}j\omega M_{1i})=V_1$. Thus, $I_1={\mathcal A} {\hat H_r}$, where
\begin{align}
{\mathcal A}=\frac{-j\omega\pi a^2 \mu_0}{R_c+j\omega L_c+1/j\omega C+\sum_{i=1,i\neq n}^{N}j\omega M_{in}},
\end{align}
which is a constant. Since the coils are uniformly distributed, $\sum_{i=1,i\neq n}^{N}j\omega M_{in}$ are the same for all the coils no matter which $n$ we choose.
Then, we consider the source is a magnetic loop antenna and ${{\bf Z}}{\hat {\bf I}}={\hat {\bf V}}$. According to \eqref{equ:voltage}, $V_n=V_0 \cos\theta_n$. Hence ${\hat{\bf V}}={\bf K}{\bf V}_0$, where ${\hat {\bf K}}$ is a diagonal matrix and $diag({\hat K_n})=\cos\theta_n$. Thus, ${{\bf Z}}{\hat {\bf I}}={\bf K}{\bf V}_0$. By multiplying the inverse of ${\bf K}$, we can obtain ${\bf K}^{-1}{ {\bf Z}}{\hat {\bf I}}={\bf V}_0$. Also, ${{\bf Z}}$ can be regarded as diagonal matrix since mutual inductances are much smaller than coil impedances, i.e., $j\omega M<<(R_c+j\omega L_c-j/\omega C)$. Thus, ${{\bf Z}}{\bf K}^{-1}{\hat {\bf I}}={\bf V}_0$. Moreover, since ${\bf Z}{\bf I}_0={\bf V}_0$, we have ${\bf I}_0={\bf K}^{-1}{\hat {\bf I}}$. Then, we can obtain ${\hat {\bf I}}={\bf K}{\bf I}_0$. As a result, ${\hat I}_n=\cos\theta_n I_n={\mathcal A} {\hat H_r}\cos\theta_n$.

%****************************************************************************
\bibliographystyle{IEEEtran}
\bibliography{meta_shell_2,ghz}

\begin{thebibliography}{10}
\providecommand{\url}[1]{#1}
\csname url@samestyle\endcsname
\providecommand{\newblock}{\relax}
\providecommand{\bibinfo}[2]{#2}
\providecommand{\BIBentrySTDinterwordspacing}{\spaceskip=0pt\relax}
\providecommand{\BIBentryALTinterwordstretchfactor}{4}
\providecommand{\BIBentryALTinterwordspacing}{\spaceskip=\fontdimen2\font plus
\BIBentryALTinterwordstretchfactor\fontdimen3\font minus
  \fontdimen4\font\relax}
\providecommand{\BIBforeignlanguage}[2]{{%
\expandafter\ifx\csname l@#1\endcsname\relax
\typeout{** WARNING: IEEEtran.bst: No hyphenation pattern has been}%
\typeout{** loaded for the language `#1'. Using the pattern for}%
\typeout{** the default language instead.}%
\else
\language=\csname l@#1\endcsname
\fi
#2}}
\providecommand{\BIBdecl}{\relax}
\BIBdecl

\bibitem{Guo_m2ipractical}
H.~Guo and Z.~Sun, ``{M2I communication: From theoretical modeling to practical
  design},'' in \emph{2016 IEEE International Conference on Communications
  (ICC)}, May 2016, pp. 1--6.

\bibitem{Markham2012}
A.~Markham and N.~Trigoni, ``Magneto-inductive networked rescue system (miners)
  taking sensor networks underground,'' in \emph{IPSN 2012}, Beijing, China,
  April 2012.

\bibitem{Jack_MI_UG_2007}
N.~Jack and K.~Shenai, ``{Magnetic Induction IC for Wireless Communication in
  RF-Impenetrable Media},'' in \emph{IEEE Workshop on Microelectronics and
  Electron Devices (WMED '07)}, April 2007.

\bibitem{masihpour2010cooperative}
M.~Masihpour and J.~I. Agbinya, ``Cooperative relay in near field magnetic
  induction: A new technology for embedded medical communication systems,'' in
  \emph{Broadband and Biomedical Communications (IB2Com), 2010 Fifth
  International Conference on}.\hskip 1em plus 0.5em minus 0.4em\relax IEEE,
  2010, pp. 1--6.

\bibitem{silva2016design}
A.~R. Silva and M.~Moghaddam, ``Design and implementation of low-power and
  mid-range magnetic-induction-based wireless underground sensor networks,''
  \emph{IEEE Transactions on Instrumentation and Measurement}, vol.~65, no.~4,
  pp. 821--835, 2016.

\bibitem{Sun_MI_TAP_2010}
Z.~Sun and I.~F. Akyildiz, ``Magnetic induction communications for wireless
  underground sensor networks,'' \emph{IEEE Transactions on Antenna and
  Propagation}, vol.~58, no.~7, pp. 2426--2435, July 2010.

\bibitem{Al-Shammaa2004}
A.~Al-Shamma'a, A.~Shaw, and S.~Saman, ``{Propagation of electromagnetic waves
  at MHz frequencies through seawater},'' \emph{Antennas and Propagation, IEEE
  Transactions on}, vol.~52, no.~11, pp. 2843--2849, Nov 2004.

\bibitem{kisseleff2014throughput}
S.~Kisseleff, I.~F. Akyildiz, and W.~H. Gerstacker, ``Throughput of the
  magnetic induction based wireless underground sensor networks: Key
  optimization techniques,'' \emph{Communications, IEEE Transactions on},
  vol.~62, no.~12, pp. 4426--4439, 2014.

\bibitem{callaham1981submarine}
M.~Callaham, ``Submarine communications,'' \emph{IEEE Communications Magazine},
  vol.~19, no.~6, pp. 16--25, 1981.

\bibitem{gulbahar2012communication}
B.~Gulbahar and O.~B. Akan, ``A communication theoretical modeling and analysis
  of underwater magneto-inductive wireless channels,'' \emph{Wireless
  Communications, IEEE Transactions on}, vol.~11, no.~9, pp. 3326--3334, 2012.

\bibitem{Domingo2012}
M.~C. Domingo, ``Magnetic induction for underwater wireless communication
  networks,'' \emph{IEEE Transactions on Antennas and Propagation}, vol.~60,
  no.~6, pp. 2929--2939, 2012.

\bibitem{through-earth-communication_2010}
L.~Yan, J.~Waynert, and C.~Sunderman, ``Measurements and modeling of
  through-the-earth communications for coal mines,'' in \emph{Industry
  Applications Society Annual Meeting (IAS), 2012 IEEE}.\hskip 1em plus 0.5em
  minus 0.4em\relax IEEE, 2012, pp. 1--6.

\bibitem{Lockheed_MI_2012}
\BIBentryALTinterwordspacing
(2010) Magnelink magnetic communication system. [Online]. Available:
  \url{http://www.teslasociety.ch/info/magnetlink/2.pdf}
\BIBentrySTDinterwordspacing

\bibitem{Michael_NIOSH_TTE_2012}
M.~R. Yenchek, G.~T. Homce, N.~W. Damiano, and J.~R. Srednicki,
  ``Niosh-sponsored research in through-the-earth communications for mines: a
  status report,'' \emph{IEEE Transactions on Industry Applications}, vol.~48,
  no.~5, pp. 1700--1707, 2012.

\bibitem{karlsson2004physical}
A.~Karlsson, ``Physical limitations of antennas in a lossy medium,''
  \emph{Antennas and Propagation, IEEE Transactions on}, vol.~52, no.~8, pp.
  2027--2033, 2004.

\bibitem{guo2015m2i}
H.~Guo, Z.~Sun, J.~Sun, and N.~Litchinitser, ``{M2I: Channel Modeling for
  Metamaterial-Enhanced Magnetic Induction Communications},'' \emph{Antennas
  and Propagation, IEEE Transactions on}, vol.~63, no.~12, pp. 1--1, December
  2015.

\bibitem{pendry1999magnetism}
J.~B. Pendry, A.~J. Holden, D.~Robbins, and W.~Stewart, ``Magnetism from
  conductors and enhanced nonlinear phenomena,'' \emph{Microwave Theory and
  Techniques, IEEE Transactions on}, vol.~47, no.~11, pp. 2075--2084, 1999.

\bibitem{scarborough2012experimental}
C.~Scarborough, Z.~Jiang, D.~Werner, C.~Rivero-Baleine, and C.~Drake,
  ``Experimental demonstration of an isotropic metamaterial super lens with
  negative unity permeability at 8.5 mhz,'' \emph{Applied Physics Letters},
  vol. 101, no.~1, p. 014101, 2012.

\bibitem{xie2012proposal}
Y.~Xie, J.~Jiang, and S.~He, ``Proposal of cylindrical rolled-up metamaterial
  lenses for magnetic resonance imaging application and preliminary
  experimental demonstration,'' \emph{Progress In Electromagnetics Research},
  vol. 124, pp. 151--162, 2012.

\bibitem{pendry2006controlling}
J.~B. Pendry, D.~Schurig, and D.~R. Smith, ``Controlling electromagnetic
  fields,'' \emph{science}, vol. 312, no. 5781, pp. 1780--1782, 2006.

\bibitem{schurig2006metamaterial}
D.~Schurig, J.~Mock, B.~Justice, S.~A. Cummer, J.~B. Pendry, A.~Starr, and
  D.~Smith, ``Metamaterial electromagnetic cloak at microwave frequencies,''
  \emph{Science}, vol. 314, no. 5801, pp. 977--980, 2006.

\bibitem{EMT_theory}
W.~Merrill, R.~Diaz, M.~Lore, M.~Squires, and N.~Alexopoulos, ``Effective
  medium theories for artificial materials composed of multiple sizes of
  spherical inclusions in a host continuum,'' \emph{IEEE Transactions on
  Antennas and Propagation}, vol.~47, no.~1, January 1999.

\bibitem{wait1951magnetic}
J.~R. Wait, ``The magnetic dipole over the horizontally stratified earth,''
  \emph{Canadian Journal of Physics}, vol.~29, no.~6, pp. 577--592, 1951.

\bibitem{durkin1984electro}
J.~Durkin, ``Electro magnatic detection of trapped miners,'' \emph{IEEE
  Communications Magazine}, vol.~22, no.~2, pp. 37--46, 1984.

\bibitem{gibson2010channel}
D.~Gibson, \emph{Channel characterisation and system design for sub-surface
  communications}.\hskip 1em plus 0.5em minus 0.4em\relax Lulu. com, 2010.

\bibitem{Akyildiz2002}
I.~F. Akyildiz, W.~Su, Y.~Sankarasubramaniam, and E.~Cayirci, ``Wireless sensor
  networks: A survey,'' \emph{Computer Networks (Elsevier) Journal}, vol.~38,
  no.~4, pp. 393--422, March 2002.

\bibitem{Shamonina2002}
E.~Shamonina, V.~A. Kalinin, K.~H. Ringhofer, and L.~Solymar,
  ``Magneto-inductive waveguide,'' \emph{Electronics Letters}, vol.~38, no.~8,
  pp. 371--373, 2002.

\bibitem{5445008}
C.~J. Stevens, C.~W.~T. Chan, K.~Stamatis, and D.~J. Edwards, ``Magnetic
  metamaterials as 1-d data transfer channels: An application for
  magneto-inductive waves,'' \emph{IEEE Transactions on Microwave Theory and
  Techniques}, vol.~58, no.~5, pp. 1248--1256, May 2010.

\bibitem{sun2011dynamic}
Z.~Sun, I.~F. Akyildiz, and G.~P. Hancke, ``Dynamic connectivity in wireless
  underground sensor networks,'' \emph{Wireless Communications, IEEE
  Transactions on}, vol.~10, no.~12, pp. 4334--4344, 2011.

\bibitem{kisseleff2015digital}
S.~Kisseleff, I.~F. Akyildiz, and W.~H. Gerstacker, ``Digital signal
  transmission in magnetic induction based wireless underground sensor
  networks,'' \emph{IEEE Transactions on Communications}, vol.~63, no.~6, pp.
  2300--2311, 2015.

\bibitem{best2007study}
S.~R. Best, ``A study of the performance properties of small antennas,'' in
  \emph{Antenna Applications Symposium}, 2007, pp. 193--219.

\bibitem{hui2001input}
H.~Hui, K.~Chan, and E.~Yung, ``The input impedance and the antenna gain of the
  spherical helical antenna,'' \emph{IEEE Transactions on Antennas and
  Propagation}, vol.~49, no.~8, pp. 1235--1237, 2001.

\bibitem{kim2010electrically}
O.~S. Kim, O.~Breinbjerg, and A.~D. Yaghjian, ``Electrically small magnetic
  dipole antennas with quality factors approaching the chu lower bound,''
  \emph{IEEE Transactions on Antennas and Propagation}, vol.~58, no.~6, pp.
  1898--1906, 2010.

\bibitem{guo2017multiple}
H.~Guo, Z.~Sun, and P.~Wang, ``Multiple frequency band channel modeling and
  analysis for magnetic induction communication in practical underwater
  environments,'' \emph{IEEE Transactions on Vehicular Technology}, 2017.

\bibitem{Sun2011a}
Z.~Sun, P.~Wang, M.~C. Vuran, M.~A. Al-Rodhaan, A.~M. Al-Dhelaan, and I.~F.
  Akyildiz, ``Mise-pipe: Magnetic induction-based wireless sensor networks for
  underground pipeline monitoring,'' \emph{Ad Hoc Networks Journal (Elsevier)},
  vol.~9, no.~3, pp. 218--227, May 2011.

\bibitem{Guo2014}
H.~Guo and Z.~Sun, ``Channel and energy modeling for self-contained wireless
  sensor networks in oil reservoirs,'' \emph{IEEE Transactions on Wireless
  Communications}, vol.~13, no.~4, pp. 2258--2269, April 2014.

\bibitem{Ziolkowski_electric}
R.~W. Ziolkowski and A.~D. Kipple, ``Application of double negative materials
  to increase the power radiated by electrically small antennas,'' \emph{IEEE
  Transactiions on Antenna and Propagation}, vol.~51, no.~10, pp. 2626--2640,
  2003.

\bibitem{lipworth2014magnetic}
G.~Lipworth, J.~Ensworth, K.~Seetharam, D.~Huang, J.~S. Lee, P.~Schmalenberg,
  T.~Nomura, M.~S. Reynolds, D.~R. Smith, and Y.~Urzhumov, ``Magnetic
  metamaterial superlens for increased range wireless power transfer,''
  \emph{Scientific reports}, vol.~4, 2014.

\bibitem{Urzhumov2011}
Y.~Urzhumov and D.~R. Smith, ``Metamaterial-enhanced coupling between magnetic
  dipoles for efficient wireless power transfer,'' \emph{Physical Review B},
  vol.~83, no.~20, 2011.

\bibitem{Scarborough2012}
C.~Scarborough, Z.~Jiang, D.~Werner, C.~Rivero-Baleine, and C.~Drake,
  ``Experimental demonstration of an isotropic metamaterial super lens with
  negative unity permeability at 8.5 mhz,'' \emph{Applied Physics Letters},
  vol. 101, no.~1, p. 014101, 2012.

\bibitem{Wang2013}
B.~Wang, W.~Yerazunis, and K.~H. Teo, ``Wireless power transfer: metamaterials
  and array of coupled resonators,'' \emph{Proceedings of the IEEE}, vol. 101,
  no.~6, 2013.

\bibitem{ranaweera2014experimental}
A.~Ranaweera, T.~P. Duong, and J.-W. Lee, ``Experimental investigation of
  compact metamaterial for high efficiency mid-range wireless power transfer
  applications,'' \emph{Journal of Applied Physics}, vol. 116, no.~4, p.
  043914, 2014.

\bibitem{zhang2015spatially}
Y.~Zhang, C.~Yao, H.~Tang, and Y.~Li, ``Spatially mapped metamaterials make a
  new magnetic concentrator for the two-coil system,'' \emph{Progress In
  Electromagnetics Research}, vol. 150, pp. 49--57, 2015.

\bibitem{yeap2016metamaterial}
S.~B. Yeap, X.~Qing, and Z.~N. Chen, ``Metamaterial magneto inductive lens for
  magnetic resonance imaging,'' in \emph{Antenna Technology (iWAT), 2016
  International Workshop on}.\hskip 1em plus 0.5em minus 0.4em\relax IEEE,
  2016, pp. 138--141.

\bibitem{freire2008experimental}
M.~J. Freire, R.~Marques, and L.~Jelinek, ``Experimental demonstration of a
  $\mu$=- 1 metamaterial lens for magnetic resonance imaging,'' \emph{Applied
  Physics Letters}, vol.~93, no.~23, p. 231108, 2008.

\bibitem{ziolkowski2006metamaterial}
R.~W. Ziolkowski and A.~Erentok, ``Metamaterial-based efficient electrically
  small antennas,'' \emph{Antennas and Propagation, IEEE Transactions on},
  vol.~54, no.~7, pp. 2113--2130, 2006.

\bibitem{erentok2008metamaterial}
A.~Erentok and R.~W. Ziolkowski, ``Metamaterial-inspired efficient electrically
  small antennas,'' \emph{Antennas and Propagation, IEEE Transactions on},
  vol.~56, no.~3, pp. 691--707, 2008.

\bibitem{palandoken2009broadband}
M.~Palandoken, A.~Grede, and H.~Henke, ``Broadband microstrip antenna with
  left-handed metamaterials,'' \emph{IEEE Transactions on Antennas and
  Propagation}, vol.~57, no.~2, pp. 331--338, 2009.

\bibitem{bait2010electromagnetic}
M.~M. Bait-Suwailam, M.~S. Boybay, and O.~M. Ramahi, ``Electromagnetic coupling
  reduction in high-profile monopole antennas using single-negative magnetic
  metamaterials for mimo applications,'' \emph{IEEE transactions on Antennas
  and Propagation}, vol.~58, no.~9, pp. 2894--2902, 2010.

\bibitem{zedler2008systematic}
M.~Zedler, ``Systematic topological design of metamaterials,'' Ph.D.
  dissertation, Ph. D. dissertation, Munich University of Technology,
  2008.[Online]. Available: http://mediatum2. ub. tum. de/node, 2008.

\bibitem{caloz2005electromagnetic}
C.~Caloz and T.~Itoh, \emph{Electromagnetic metamaterials: transmission line
  theory and microwave applications}.\hskip 1em plus 0.5em minus 0.4em\relax
  John Wiley \& Sons, 2005.

\bibitem{horn1984extended}
B.~K. Horn, ``Extended gaussian images,'' \emph{Proceedings of the IEEE},
  vol.~72, no.~12, pp. 1671--1686, 1984.

\bibitem{mohan1999simple}
S.~S. Mohan, M.~del Mar~Hershenson, S.~P. Boyd, and T.~H. Lee, ``Simple
  accurate expressions for planar spiral inductances,'' \emph{Solid-State
  Circuits, IEEE Journal of}, vol.~34, no.~10, pp. 1419--1424, 1999.

\bibitem{jow2007design}
U.-M. Jow and M.~Ghovanloo, ``Design and optimization of printed spiral coils
  for efficient transcutaneous inductive power transmission,'' \emph{Biomedical
  Circuits and Systems, IEEE Transactions on}, vol.~1, no.~3, pp. 193--202,
  2007.

\bibitem{wheeler1942formulas}
H.~A. Wheeler, ``Formulas for the skin effect,'' \emph{Proceedings of the IRE},
  vol.~30, no.~9, pp. 412--424, 1942.

\bibitem{Guo_underwater}
H.~Guo, Z.~Sun, and P.~Wang, ``Channel modeling of mi underwater communication
  using tri-directional coil antenna,'' in \emph{IEEE Globecom 2015}, San
  Diego, USA, Dec 2015.

\bibitem{Ziolkowski_antenna_efficiency_2007}
R.~W. Ziolkowski and A.~Erentok, ``At and below the chu limit: passive and
  active broad bandwidth metamaterial-based electrically small antennas.''
  \emph{IET microwaves, antennas, and propagation}, vol.~1, no.~1, pp.
  116--128, 2007.

\bibitem{tai1947hertzian}
C.~Tai, ``Hertzian dipole immersed in a dissipative medium: Cruft. lab,''
  Report, Tech. Rep., 1947.

\bibitem{wait1952magnetic}
J.~Wait, ``The magnetic dipole antenna immersed in a conducting medium,''
  \emph{Proceedings of the IRE}, vol.~10, no.~40, pp. 1244--1245, 1952.

\bibitem{james1957insulated}
R.~James, ``Insulated loop antenna immersed in a conducting medium,''
  \emph{Journal of Research of the National Bureau of Standards}, vol.~59,
  no.~2, 1957.

\bibitem{large1973radio}
D.~Large, L.~Ball, and A.~Farstad, ``Radio transmission to and from underground
  coal mines--theory and measurement,'' \emph{IEEE Transactions on
  Communications}, vol.~21, no.~3, pp. 194--202, 1973.

\bibitem{Balanis_a}
C.~A. Balanis, \emph{Antenna Theory}.\hskip 1em plus 0.5em minus 0.4em\relax
  John Wiley Publishing Company, 2005.

\bibitem{rappaport1996wireless}
T.~S. Rappaport \emph{et~al.}, \emph{Wireless communications: principles and
  practice}.\hskip 1em plus 0.5em minus 0.4em\relax prentice hall PTR New
  Jersey, 1996, vol.~2.

\bibitem{pittman1985through}
W.~E. Pittman, R.~H. Church, and J.~T. McLendon, \emph{Through-the-earth
  electromagnetic trapped miner location systems: A review}.\hskip 1em plus
  0.5em minus 0.4em\relax US Department of Interior, Bureau of Mines, 1985.

\bibitem{wiltshire2014noise}
M.~Wiltshire and R.~Syms, ``Noise performance of magneto-inductive cables,''
  \emph{Journal of Applied Physics}, vol. 116, no.~3, p. 034503, 2014.

\bibitem{stevens2010magnetic}
C.~J. Stevens, C.~W. Chan, K.~Stamatis, and D.~J. Edwards, ``Magnetic
  metamaterials as 1-d data transfer channels: an application for
  magneto-inductive waves,'' \emph{IEEE Transactions on Microwave Theory and
  Techniques}, vol.~58, no.~5, pp. 1248--1256, 2010.

\bibitem{pozar2009microwave}
D.~M. Pozar, \emph{Microwave engineering}.\hskip 1em plus 0.5em minus
  0.4em\relax John Wiley \& Sons, 2009.

\bibitem{lipworth2015quasi}
G.~Lipworth, J.~Ensworth, K.~Seetharam, J.~S. Lee, P.~Schmalenberg, T.~Nomura,
  M.~S. Reynolds, D.~R. Smith, and Y.~Urzhumov, ``Quasi-static magnetic field
  shielding using longitudinal mu-near-zero metamaterials,'' \emph{Scientific
  reports}, vol.~5, 2015.

\end{thebibliography}

\end{document}